%% file: draft.tex
\documentclass[aps,prd, twocolumn, showpacs,amsmath,amssymb]{revtex4-1}
\usepackage{graphicx}
\usepackage{dcolumn}
\usepackage{bm}
\usepackage{epsfig}
\usepackage{rotating}
\usepackage{color}

\newcommand{\eff}{\varepsilon}
\newcommand{\psip}{\psi(3686)}

\newcommand{\jpsi}{J/\psi}
\newcommand{\chicJ}{\chi_{cJ}}

\newcommand{\chico}{\chi_{c1}}
\newcommand{\chict}{\chi_{c2}}
\newcommand{\EE}{e^+e^-}

\newcommand{\pip}{\pi^+}
\newcommand{\pim}{\pi^-}
\newcommand{\pio}{\pi^0}

\newcommand{\bfg}{\begin{figure}}
\newcommand{\efg}{\end{figure}}
\newcommand{\bitm}{\begin{itemize}}
\newcommand{\eitm}{\end{itemize}}
\newcommand{\bnum}{\begin{enumerate}}
\newcommand{\enum}{\end{enumerate}}
\newcommand{\btbl}{\begin{table}}
\newcommand{\etbl}{\end{table}}
\newcommand{\btbu}{\begin{tabular}}
\newcommand{\etbu}{\end{tabular}}

\newcommand{\KKB}{K\bar{K}}
\newcommand{\QQB}{q\bar{q}}

\newcommand{\beq}{\begin{equation}}
\newcommand{\edq}{\end{equation}}

\newcommand{\jpc}{J^{PC}}

\newcommand{\etap}{\eta^{\prime}}

\newcommand{\fzero}{f_{0}(980)}
\newcommand{\azero}{a_{0}(980)}
\newcommand{\azeAmp}{\azero\pi}
\newcommand{\atwo}{a_{2}(1320)}
\newcommand{\atwoAmp}{\atwo\pi}
\newcommand{\atwop}{a_{2}(1700)}
\newcommand{\atwopAmp}{\atwop\pi}
\newcommand{\ftwo}{f_{2}(1270)}
\newcommand{\ftwoAmp}{\ftwo\eta}
\newcommand{\ffo}{f_{4}(2050)}
\newcommand{\ffoAmp}{\ffo\eta}
\newcommand{\pionef}{\pi_{1}(1400)}

\newcommand{\piones}{\pi_{1}(1600)}

\newcommand{\spp}{S_{\pi\pi}}
\newcommand{\skk}{S_{\KKB}}
\newcommand{\sppz}{S^{0}_{\pi\pi}}
\newcommand{\szs}{S^{0}(s)}

\newcommand{\skkAmp}{\skk\eta}
\newcommand{\sppAmp}{\spp\eta}
\newcommand{\decaya}{\chico\to\eta\pip\pim}
\newcommand{\decayb}{\chico\to\etap\pip\pim}
\newcommand{\fulldecaya}{\psip\to\gamma\chico;~\decaya}

\newcommand{\cc}{c^2}
\newcommand{\gev}{\mathrm{GeV}}

\newcommand{\gevcc}{\mathrm{GeV}/\cc}
\newcommand{\mevcc}{\mathrm{MeV}/\cc}

\begin{document}
\normalsize
\parskip=5pt plus 1pt minus 1pt

\title{\boldmath 
Amplitude analysis of the $\chi_{c1} \to \eta\pi^+\pi^-$ decays} 

\input{authors_PUB153}

\date{\today}

\begin{abstract}

 Using $448.0 \times 10^6$~$\psi(3686)$ events collected with the BESIII detector, an
 amplitude analysis is performed for $\psi(3686)\to\gamma\chi_{c1}$, $\chi_{c1}\to\eta\pi^+\pi^-$
 decays. The most dominant two-body structure observed is
 $a_0(980)^{\pm}\pi^{\mp}$; $a_0(980)^{\pm}\to\eta\pi^{\pm}$.  The
 $a_0(980)$ line shape is modeled using a dispersion relation, and a
 significant nonzero $a_0(980)$ coupling to the $\eta^{\prime}\pi$ channel is
 measured. We observe $\chi_{c1}\to a_2(1700)\pi$ production for the
 first time, with a significance larger than 17$\sigma$.  The
 production of mesons with exotic quantum numbers, $J^{PC}=1^{-+}$, is
 investigated, and upper limits for the branching fractions $\chi_{c1}\to
 \pi_1(1400)^{\pm}\pi^{\mp}$, $\chi_{c1}\to \pi_1(1600)^{\pm}\pi^{\mp}$,
 and $\chi_{c1}\to \pi_1(2015)^{\pm}\pi^{\mp}$, with subsequent
 $\pi_1(X)^{\pm} \to \eta\pi^{\pm}$ decay, are determined.

\end{abstract}


\maketitle

\section{Introduction}\label{Intro}

Charmonium decays provide a rich laboratory for light meson
spectroscopy. Large samples of charmonium states with $\jpc=1^{--}$,
like the $\jpsi$ and $\psip$, are easily produced at $\EE$ colliders,
and their transitions provide sizable charmonium samples with other
$\jpc$ quantum numbers, like the $\chico$ ($1^{++}$).  The
$\chico\to\eta\pi\pi$ decay is suitable for studying the production of
exotic mesons with $\jpc = 1^{-+}$, which could be observed decaying
into the $\eta\pi$ final state. The lowest orbital excitation of a
two-body combination in $\chico$ decays to three pseudoscalars, for
instance $\chico\to\eta\pi\pi$, is the $S$-wave transition, in which
if a resonance is produced, it has to have $\jpc = 1^{-+}$.  Several
candidates with $\jpc=1^{-+}$, decaying into different final states, 
such as $\eta\pi$, $\etap\pi$, $f_1(1270)\pi$, $b_1(1235)\pi$ and
$\rho\pi$, have been reported by various experiments, and these have
been thoroughly reviewed in Ref.~\cite{CMeyer}. The lightest exotic
meson candidate is the $\pionef$~\cite{PDG14}, reported only in the
$\eta\pi$ final state 
by GAMS~\cite{GAMS88}, KEK~\cite{Aoyagi}, Crystal Barrel~\cite{CBARpi1}, and E852~\cite{E852pi1}, 
but its resonance nature is controversial~\cite{DONADAM}. The most promising
$\jpc = 1^{-+}$ candidate, the $\pi_1(1600)$~\cite{PDG14}, could also
couple to the $\eta\pi$, since it has been observed in the $\etap\pi$
channel 
by VES~\cite{Pi1Etap} and E852~\cite{Pi1E852}.

The CLEO-c collaboration reported evidence of an exotic signal in
$\decayb$ decays, consistent with $\piones\to\etap\pi$
production~\cite{CLEOMK}. However, other possible exotic signals that
could be expected have not been observed in either $\decaya$ or
$\decayb$ decays. With a more than 15 times larger data sample at BESIII,
there is an opportunity to search for the production of $\pi_1$ exotic
mesons. 
In this work we investigate possible production of exotic mesons in
the mass region 
(1.3--2.0)~GeV/$c^2$, 
decaying into the $\eta\pi^+$ +
c.c. final state, namely the $\pi_1(1400)$, $\pi_1(1600)$, and
$\pi_1(2015)$, using $\chico\to\eta\pip\pim$ decays. Charge
conjugation and isospin symmetry are assumed in this analysis.

Additional motivation for studying these decays is that a very
prominent $\azero\to\eta\pi$ signal of high purity was observed in
$\chico\to\eta\pip\pim$, by CLEO-c~\cite{CLEOMK}. The $\azero$ was discovered
several decades ago, but its nature 
was puzzling from the beginning, leading to the hypothesis that it is a four-quark rather than an ordinary 
$\QQB$ state~\cite{Jaffe77,Close93,AchasIvan89}.
The first coupled meson-meson ($\eta\pi,\KKB, \etap\pi$) scattering amplitudes 
based on lattice QCD calculations~\cite{Dudeka0} 
indicate that the $\azero$ might be a resonance strongly coupled to $\eta\pi$ and $\KKB$ channels,
which does not manifest itself as a symmetric bump in the spectra. 
Recent theoretical work based on the chiral unitarity approach also points that the $\azero$, as well as 
the $\sigma$ and $\fzero$ states, could be dynamically generated through meson-meson interactions,
for example in heavy-meson decays: $\chico\to\eta\pi\pi$~\cite{OsetChic1} and $\eta_c\to\eta\pi\pi$~\cite{OsetEtac}. 
However, there is still no consensus on the exact role that meson-meson loops play 
in forming of the $\azero$, which is now generally accepted as a four-quark object, see ~\cite{Wolk2015} and reference therein.

The $\azero$ indeed decays dominantly into $\eta\pi$ and $\KKB$ final states; the latter has a profound
influence on the $\azero$ line shape in the $\eta\pi$ channel, due to
the proximity of the $\KKB$ threshold to the $\azero$ mass. 
Different experiments, E852~\cite{Teige}, Crystal Barrel~\cite{AmslerPLB333,DBUG78_08} and 
CLEO-c~\cite{CLEOMK} analyzed data
to determine the couplings of the $\azero$ to the $\eta\pi$ ($g_{\eta\pi}$) 
and $\KKB$ final states ($g_{\KKB}$), in order to help resolve 
the true nature of the $\azero$.
This is not an exhaustive list of analyses: it points out that the values obtained 
for the $\azero$ parameters vary considerably among various 
analyses. 

Another channel of interest
is $\azero\to\etap\pi$, with the threshold more than 100~$\mevcc$
above the $\azero$ mass. The first direct observation of the decay $\azero \to
\etap\pi$ was reported by CLEO-c~\cite{CLEOMK}, using a sample
of $26\times 10^6$ $\psip$ decays. The $\azero$ coupling to the
$\etap\pi$ channel, $g_{\etap\pi}$, was determined from
$\chico\to\eta\pip\pim$ decays, although the analysis was not very
sensitive to the $\azero\to\etap\pi$ component in the
$\azero\to\eta\pi$ invariant mass distribution, and $g_{\etap\pi}$
was found to be consistent with zero. In many analyses of $\azero$
couplings, $g_{\etap\pi}$ has not been measured. For example, its
value was fixed in Ref.~\cite{DBUG78_08} based on SU(3) flavor-mixing
predictions. Using a clean sample of $\chico$ produced in the
radiative transition $\psip\to\gamma\chico$ at BESIII, we investigate
the $\chico\to\eta\pip\pim$ decays to test if the $\azero\to\eta\pi$
invariant mass distribution is sensitive to $\etap\pi$
production. Dispersion integrals in the description of the $\azero$
line shape are used to determine the $\azero$ parameters, its
invariant mass, $m_{\azero}$, and three coupling constants,
$g_{\eta\pi}$, $g_{\KKB}$ and $g_{\etap\pi}$. This information might 
help in determining the quark structure of the $\azero$.

In this $\chico$ decay mode, it is also possible to study
$\chico\to\atwop\pi$; $\atwop\to\eta\pi$ production. The $\atwop$ has
been reported 
in this decay mode by 
Crystal Barrel~\cite{a21700exp} and Belle~\cite{a21700exp_2},
but still is not accepted as an established
resonance by the Particle Data Group (PDG)~\cite{PDG14}.

\section{Event selection}\label{Data}

For our studies we use $(448.0\pm3.1)\times 10^6$ $\psip$
events, collected in 2009~\cite{BESnPsip} and 2012~\cite{Psip2012} with
the BESIII detector~\cite{bes3}.  We select 95\% of possible $\eta$
decays, in the $\eta\to\gamma\gamma$, $\eta\to\pip\pim\pio$ and
$\eta\to\pio\pio\pio$ decay modes. For each
$\psip\to\gamma\eta\pip\pim$ final state topology, exclusive Monte
Carlo (MC) samples are generated according to the relative branching
fractions given in Table~\ref{tab:fs}, equivalent to a total of
$2\times 10^7$ $\fulldecaya$ events. The background is studied using
an inclusive MC sample of 106$\times10^6$ generic $\psip$ events.

BESIII is a conventional solenoidal magnet detector that has almost
full geometrical acceptance, 
and four main components: the main drift chamber (MDC),
electromagnetic calorimeter (EMC), time-of-flight detector, all
enclosed in 1~T magnetic field, and the muon chamber. The momentum
resolution for 
majority of charged particles is better than 0.5\%.
The energy resolution for 1.0~GeV photons in the barrel (end-cap) region of the
EMC is 2.5\% (5\%). 
For the majority of photons in the barrel region, with the energy between 100 and 200 MeV, 
the energy resolution is better than 4\%. 
Details of the BESIII detector and its performance
can be found in Ref~\cite{bes3}.

Good photon candidates are selected from isolated EMC showers with energy
larger than 25 (50) MeV in the barrel (end-cap) region, corresponding
to the polar angle, $\theta$, satisfying $|\cos{\theta} | < 0.80$ ($
0.86 < | \cos{\theta} | < 0.92$). The timing of good EMC showers is
required to be within 700~ns of the trigger time. Charged tracks must
satisfy $|\cos{\theta} | < 0.93$, and the point of closest approach of
a track from the interaction point along the beam direction is
required to be within 20~cm and within 2~cm perpendicular to the beam
direction. All charged tracks are assumed to be pions, and the
inclusive MC sample is used to verify that the kaon contamination in
the final sample is negligible in each of the $\eta$ channels. We
require two charged tracks for the $\eta\to\gamma\gamma$ and $\eta\to
3\pio$ channels, and four tracks for the $\eta\to\pip\pim\pio$
channel, with zero net charge. For $\eta\to\gamma\gamma$ and
$\eta\to\pip\pim\pio$, at least three photon candidates are required, and
for $\eta\to 3\pio$ at least seven photon candidates. The invariant mass
of two-photon combinations is kinematically constrained to the $\pio$
or $\eta$ mass.

The sum of momenta of all final-state particles, for a given final
state topology, is constrained to the initial $\psip$ momentum.
If multiple combinations for an event are found, the one
with the smallest $\chi^2_{NC}$ is retained.  Here $NC$ refers to the
number of constraints, which is four plus the number of two-photon
$\pio$ and $\eta$ candidates in the final state (see
Table~\ref{tab:fs}).

\begin{table}[htbp]
\caption{
Characteristics of the $\eta$ decay channels used to reconstruct the $\psip\to\gamma\eta\pip\pim$ decays:
branching fraction $\mathcal{B}$, final state topology, number of constraints (NC) in the kinematic fit, 
and reconstruction efficiency, $\eff$, according to exclusive phase-space MC. 
\setlength{\tabcolsep}{24pt}
\label{tab:fs}
}
\begin{center}
\begin{tabular}{l@{\hskip 0.2cm}c@{\hskip 0.2cm}c@{\hskip 0.2cm}c@{\hskip 0.2cm}c} 
\hline\hline
	Decay & $\mathcal{B}$ [$\%$]~\cite{PDG14} & Final state & NC & $\eff$~[\%] \\
\hline 
$\eta\to\gamma\gamma$ & 39.41$\pm$0.20 & $3\gamma~1(\pip\pim)$ & 5 & 26.58 \\ 
$\eta\to\pip\pim\pio$ & 22.92$\pm$0.28 & $3\gamma~2(\pip\pim)$ & 5 & 16.46 \\
$\eta\to\pio\pio\pio$ & 32.68$\pm$0.23 & $7\gamma~1(\pip\pim)$ & 7 & 5.64 \\
 \hline 
Total & 95.01$\pm$0.71 & & & 16.91\\
\hline\hline
\end{tabular}
\end{center}
\end{table}

\subsection{\boldmath $\chico\to\eta\pip\pim$ event selection}\label{EvtSel} 

The $\chico\to\eta\pip\pim$ candidates in $\eta$ three-pion decays are
selected by requiring that the invariant mass of three pions satisfy
\begin{equation}\label{eq:etaM}
0.535 < m(3\pi) < 0.560~\gevcc.
\end{equation}
For the $\eta\to\gamma\gamma$ candidates, we require that the mass
constraint fit for $\eta\to\gamma\gamma$ satisfies
$\chi^2_{\gamma\gamma} < 15$. The $\chi^2_{NC}$ obtained from
four-momenta kinematic constraint fits are required to satisfy
$\chi^2_{5C} < 40 $, $\chi^2_{5C} < 40 $ and $\chi^2_{7C} < 56 $ for
$\eta\to\gamma\gamma$, $\eta\to\pip\pim\pio$ and $\eta\to3\pio$,
respectively.  
These selection criteria effectively remove kaon and other charged track contamination, 
justifying the  assumption that all charged tracks are pions. 
To select the $\chico$ candidates from the
$\psip\to\gamma\chico$ transition, we require the energy of the
radiative photon to satisfy $0.155 < E_{\gamma} < 0.185~\gev$.

\subsubsection{Background suppression}\label{BcgSup}

The major background for all final states comes from
$\psip\to\eta\jpsi$, while in the $\eta\to\gamma\gamma$ case the
background from $\psip\to\gamma\gamma\jpsi$ decays is also
significant.  The background from $\psip\to\pi\pi\jpsi$ is negligible,
once a good $\eta$ candidate is found.

To suppress the $\psip\to\eta\jpsi$ background for all three $\eta$
decays, the system recoiling against the $\eta$, with respect to the
$\psip$, must have its invariant mass separated at least $20~\mevcc$
from the $\jpsi$ mass.

Additional selection criteria are used in the $\eta\to\gamma\gamma$
channel to suppress $\pio$ contamination and
$\psip\to\gamma\gamma\jpsi$ production. The former background is
suppressed by rejecting events in which any two-photon combination
satisfies $0.110 < m(\gamma\gamma) < 0.155~\gevcc$. The latter
background is suppressed by vetoing events for which a two-photon
combination not forming an $\eta$ has a total energy between
$0.52~\gev <E_{\gamma\gamma} < 0.60~\gev$.  This range of energies is
associated with the doubly radiative decay $\psip\to\gamma\chicJ;
\chicJ\to\gamma\jpsi$, for which the energy sum of two transitional
photons is $E_{\gamma\gamma} \approx 0.560$~GeV.

\subsubsection{Background subtraction}\label{BcgSub}

\begin{figure*}[htbp]
\centerline{\hbox{ 
\epsfig{file=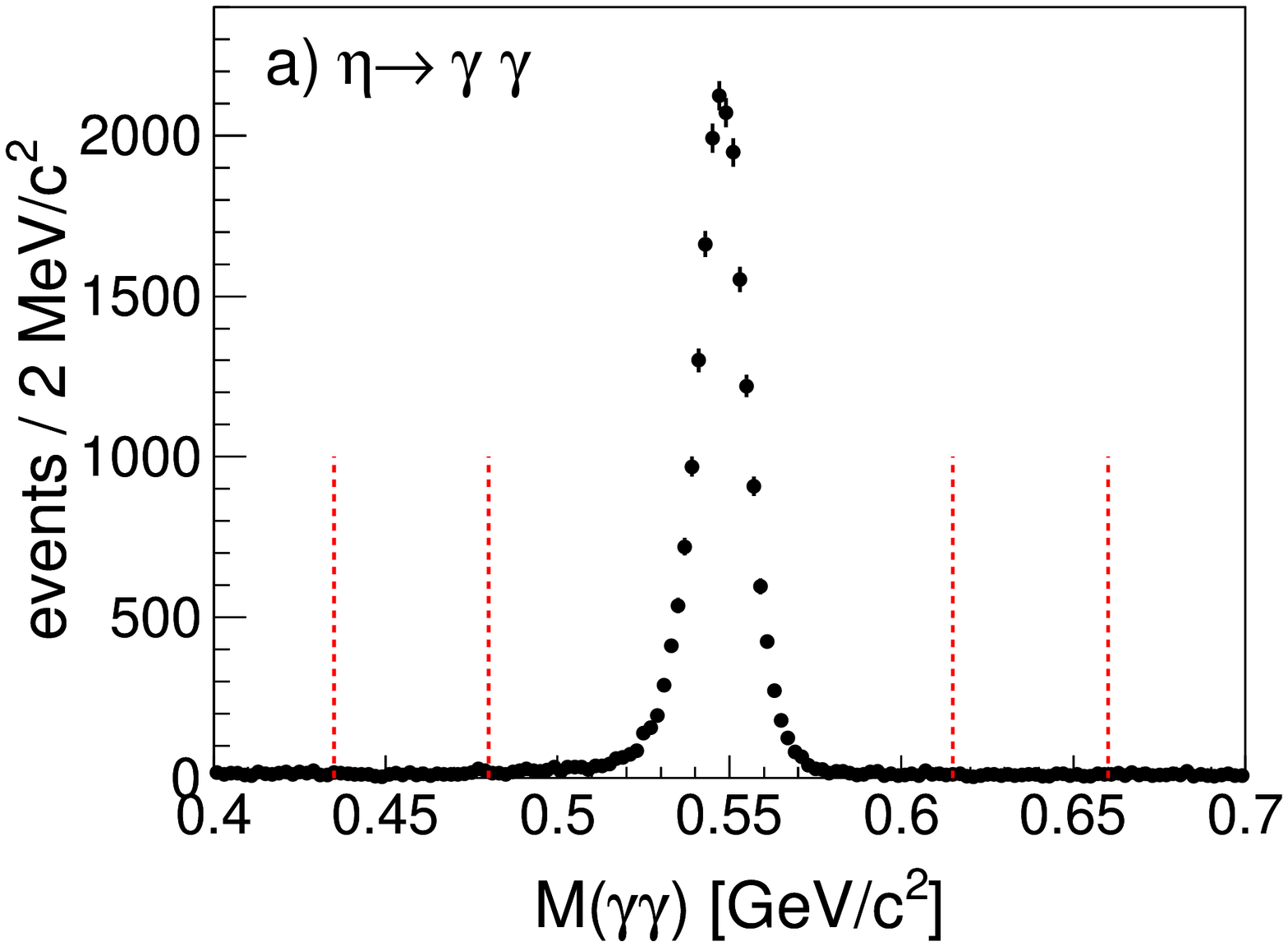,width=6.0cm} 
\epsfig{file=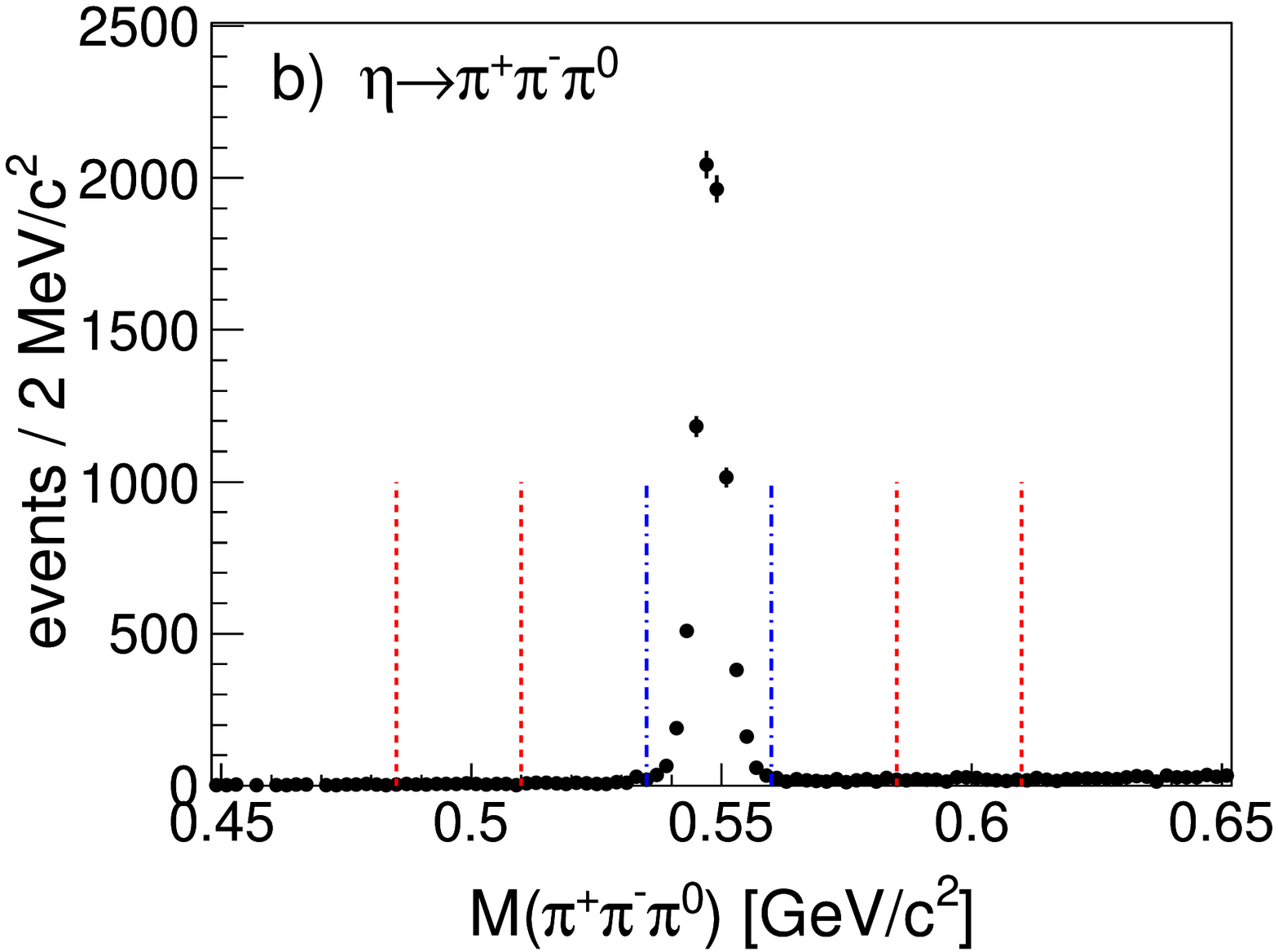,width=6.0cm} 
\epsfig{file=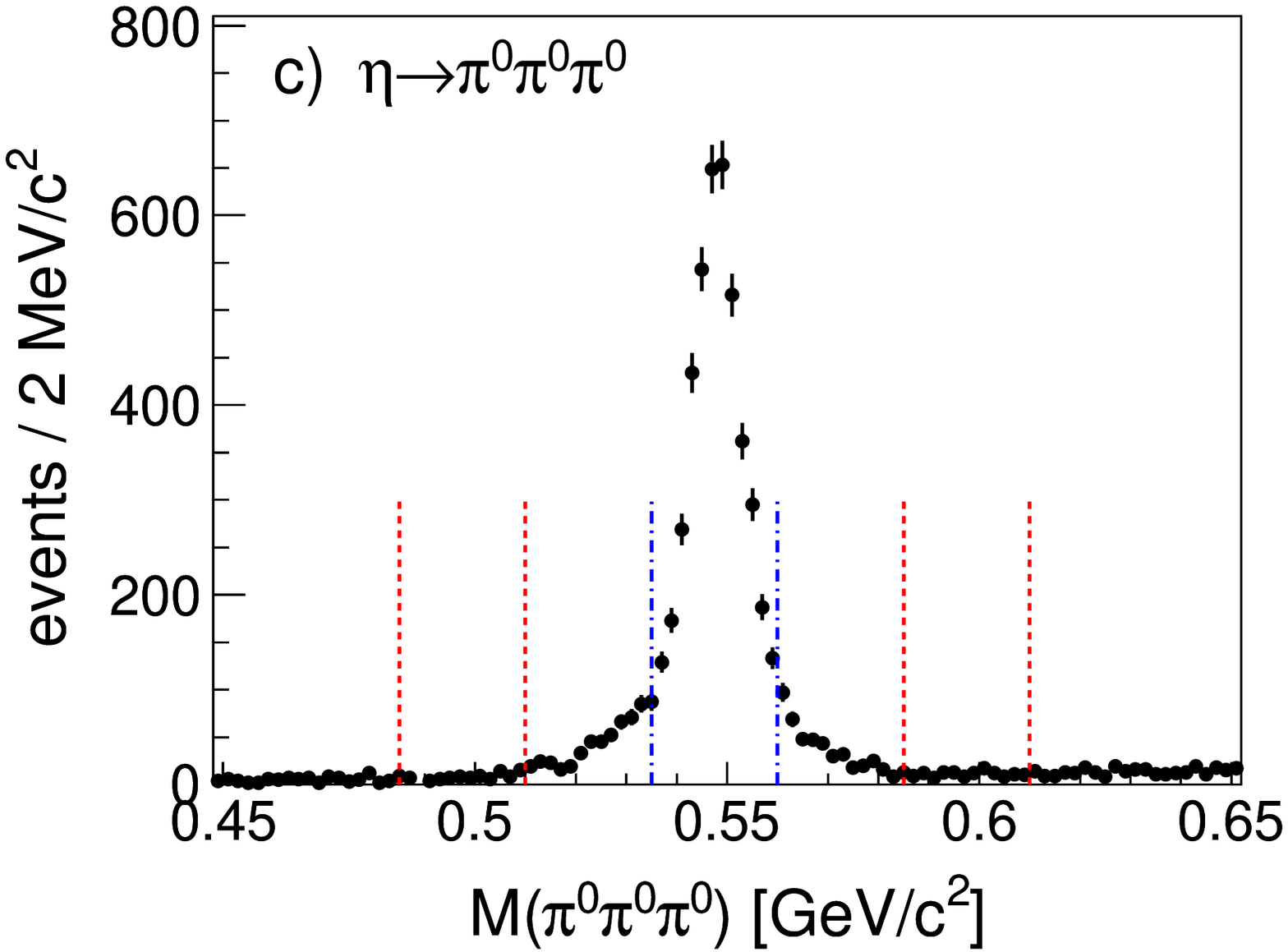,width=6.0cm}
}}

\caption{The invariant mass distribution of the $\eta$ candidates,
	where dotted (red) lines indicate regions used for background
	subtraction, while dash-dotted bars (blue) show $\eta$-signal boundaries for the
 three-pion $\eta$ decay cases. 
 There are no blue bars on plot (a) since the $\eta\to\gamma\gamma$ signal is selected 
 using the $\gamma\gamma$ kinematic constraint.
\label{fig:etaMass}
 }
\end{figure*}

The background estimated from the inclusive MC after all 
selection criteria are applied is below 3\% in each channel. The
background from $\eta$ sidebands is subtracted, and
Fig.~\ref{fig:etaMass} shows the invariant mass distributions of
$\eta$ candidates with vertical dotted bars showing the $\eta$ sideband
regions. The sideband regions for the two-photon and three-pion modes
are defined as $ 68 < | m(\gamma\gamma) - m_{\eta} | < 113~\mevcc $
and $ 37 < | m(3\pi) - m_{\eta} | < 62~\mevcc$, respectively, where
$m_{\eta}$ is the nominal $\eta$ mass~\cite{PDG14}. In the case of
$\eta$ three-pion decays, the $\eta$ signal region, defined by
Eq.~(\ref{eq:etaM}), is indicated by dash-dotted bars in
Fig.~\ref{fig:etaMass}.  Although the mass distribution of three
neutral pions, Fig.~\ref{fig:etaMass}(c), is wider than the
corresponding distribution from the charged channel,
Fig.~\ref{fig:etaMass}(b), we use the same selection criteria for both
$\eta$ decays, 
which keeps the majority of
good $\eta\to 3\pio$ candidates and results in similar background
levels in the two channels.  The effects of including more data from
the tails of these distributions are taken into account in the systematic
uncertainties.  The invariant mass plot representing $\eta\to\gamma\gamma$
candidates, Fig.~\ref{fig:etaMass}(a), is used only to select $\eta$
sidebands for background subtraction.  Table~\ref{tab:fs} lists
channel efficiencies and the effective efficiency for all channels.

\begin{figure}[htbp]
\centerline{\hbox{
\epsfig{file=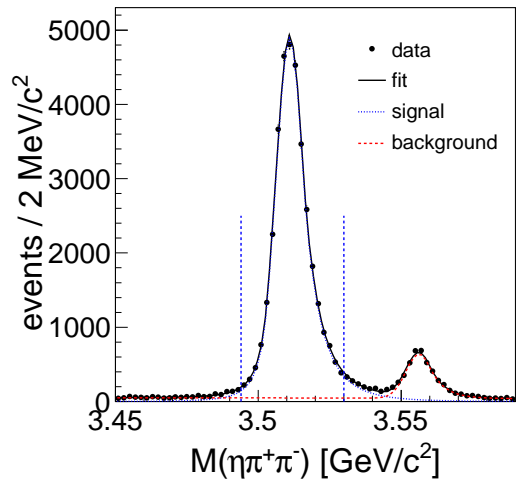,width=7.cm} 
}}
\caption{ Invariant mass of the $\chico$ candidates, after the
 $\eta$ sideband background is subtracted. Vertical bars indicate the 
 region used to select the $\chico$ candidates. See the text for the fit discussion.} \label{fig:Mchic}
\end{figure}

The $\eta\pip\pim$ invariant mass distribution, when events from all
$\eta$ channels are combined, is shown in Fig.~\ref{fig:Mchic}.  In
the signal region, indicated by vertical bars, there are 33919 events,
with the background of 497 events estimated from the $\eta$ sidebands.
The sideband background does not account for all the background, and
after the $\eta$-sideband background is subtracted, the remaining
background is estimated by fitting the invariant mass distribution. %
The fit is shown by the solid distribution, 
Fig.~\ref{fig:Mchic}.  For the $\chico$ signal, a double-sided
Crystal-Ball distribution (dotted) is used, and for the background, a
linear function along with a Gaussian corresponding to the $\chict$
contribution (dashed) are used.  The signal purity estimated from the
fit is $\mathcal{P} = (98.5\pm0.3)\%$, where the error is obtained from
fluctuations in the background when using different fitting ranges
and shapes of the background.

\subsection{\boldmath Two-body structures in the
 $\chico\to\eta\pip\pim$ decays}\label{SigAll}

\begin{figure*}[htbp]
\epsfig{file=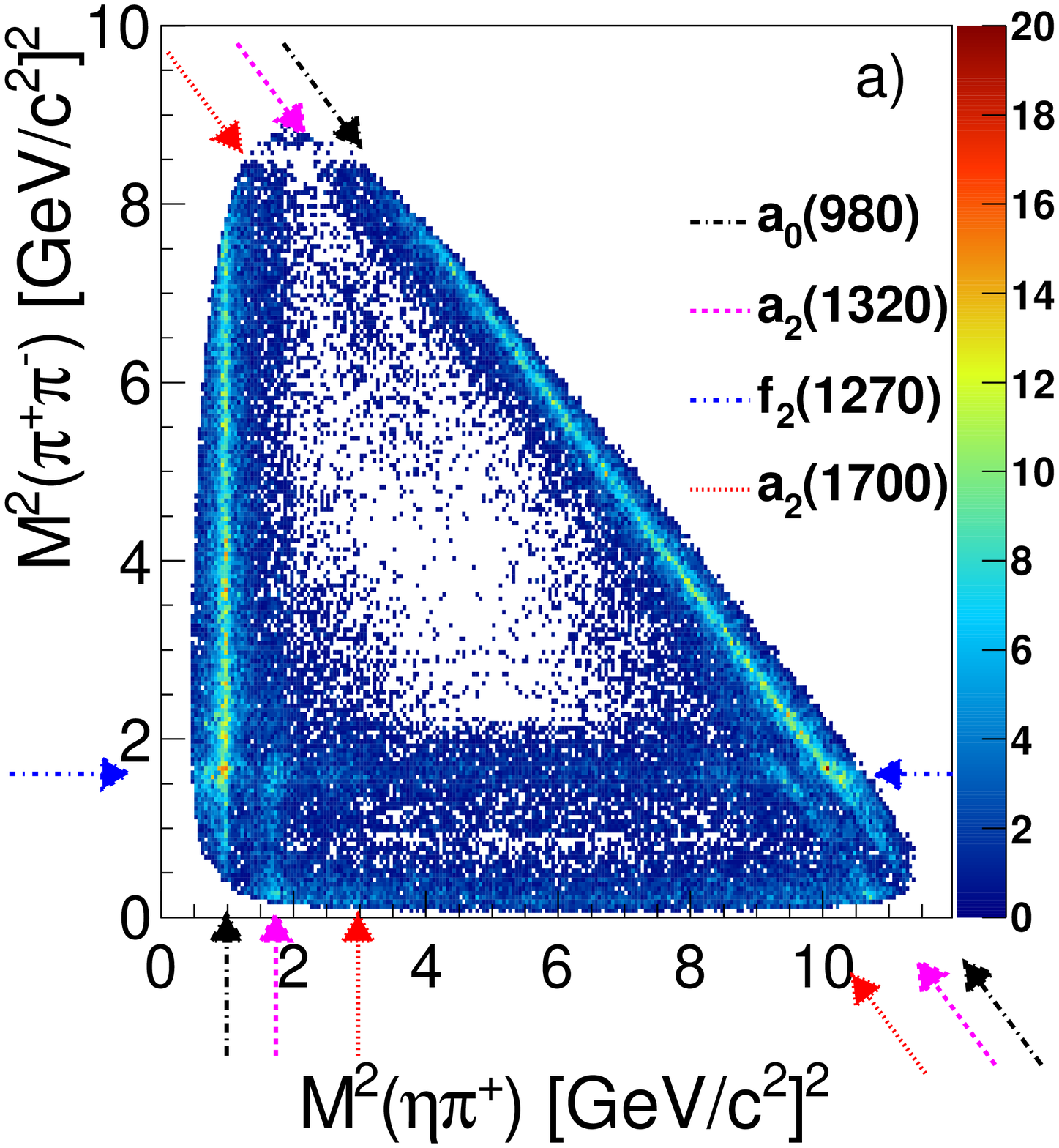,width=7.cm} 
\epsfig{file=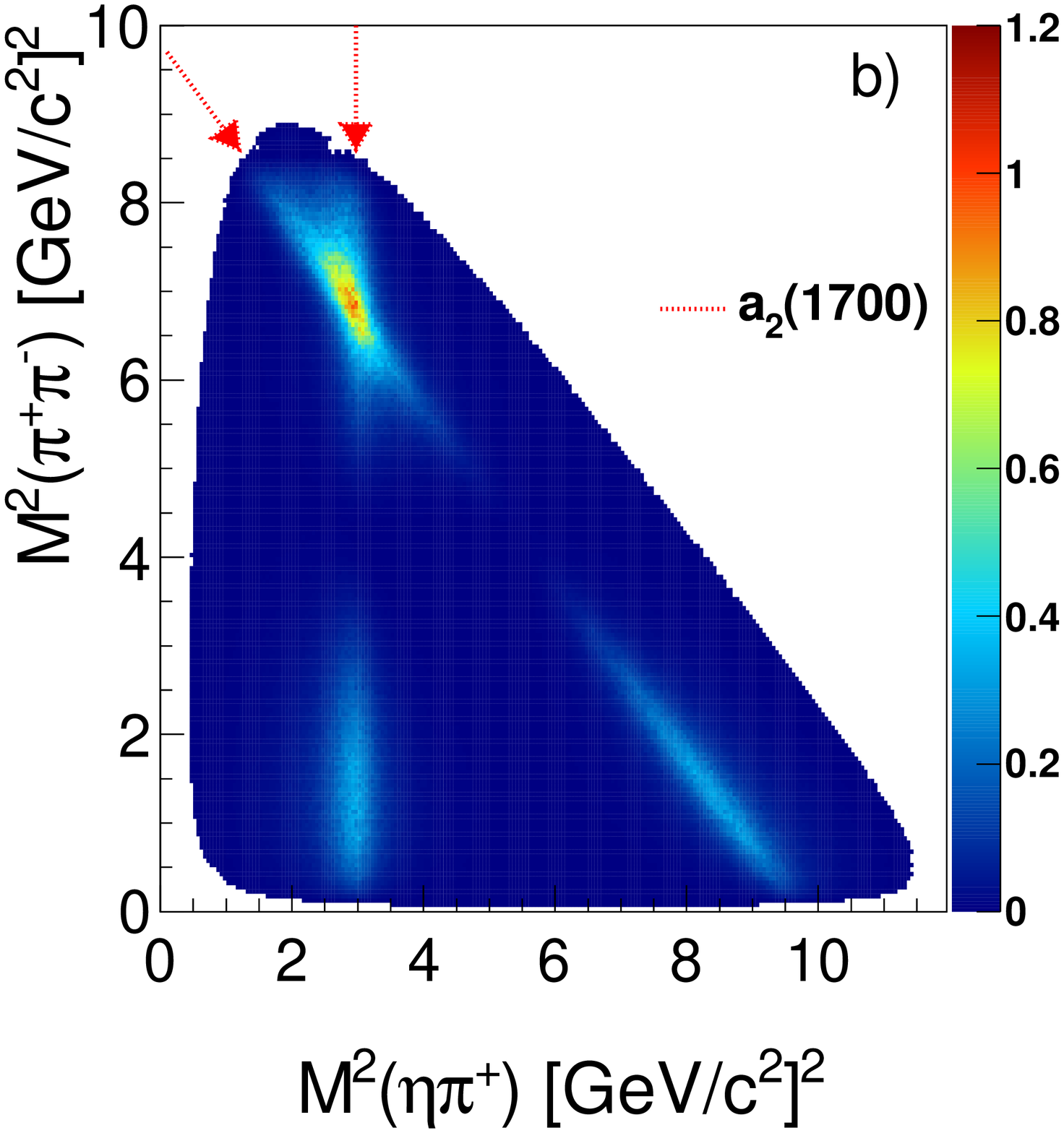,width=7.0cm} 
\epsfig{file=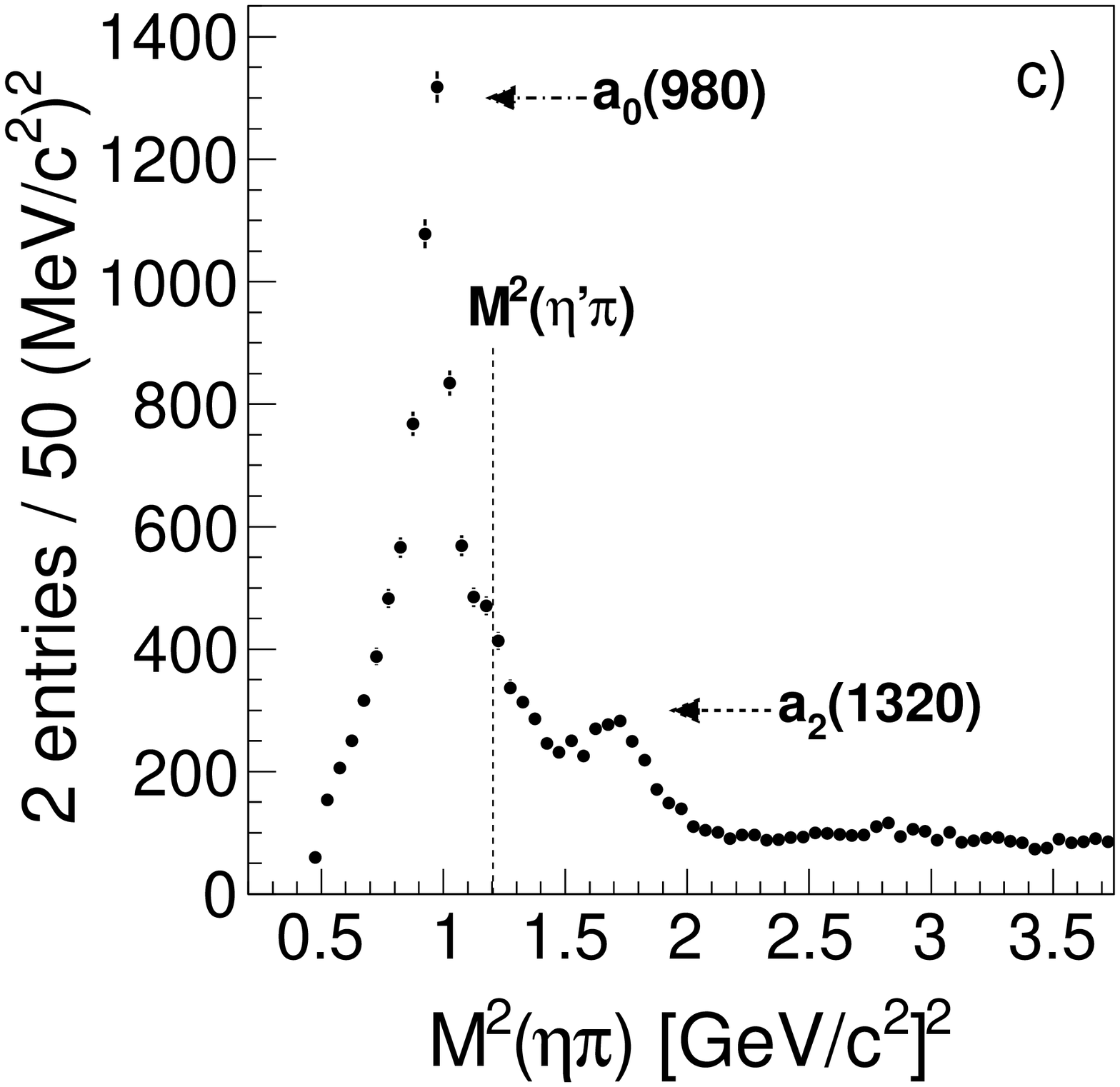,width=7.0cm} 
\epsfig{file=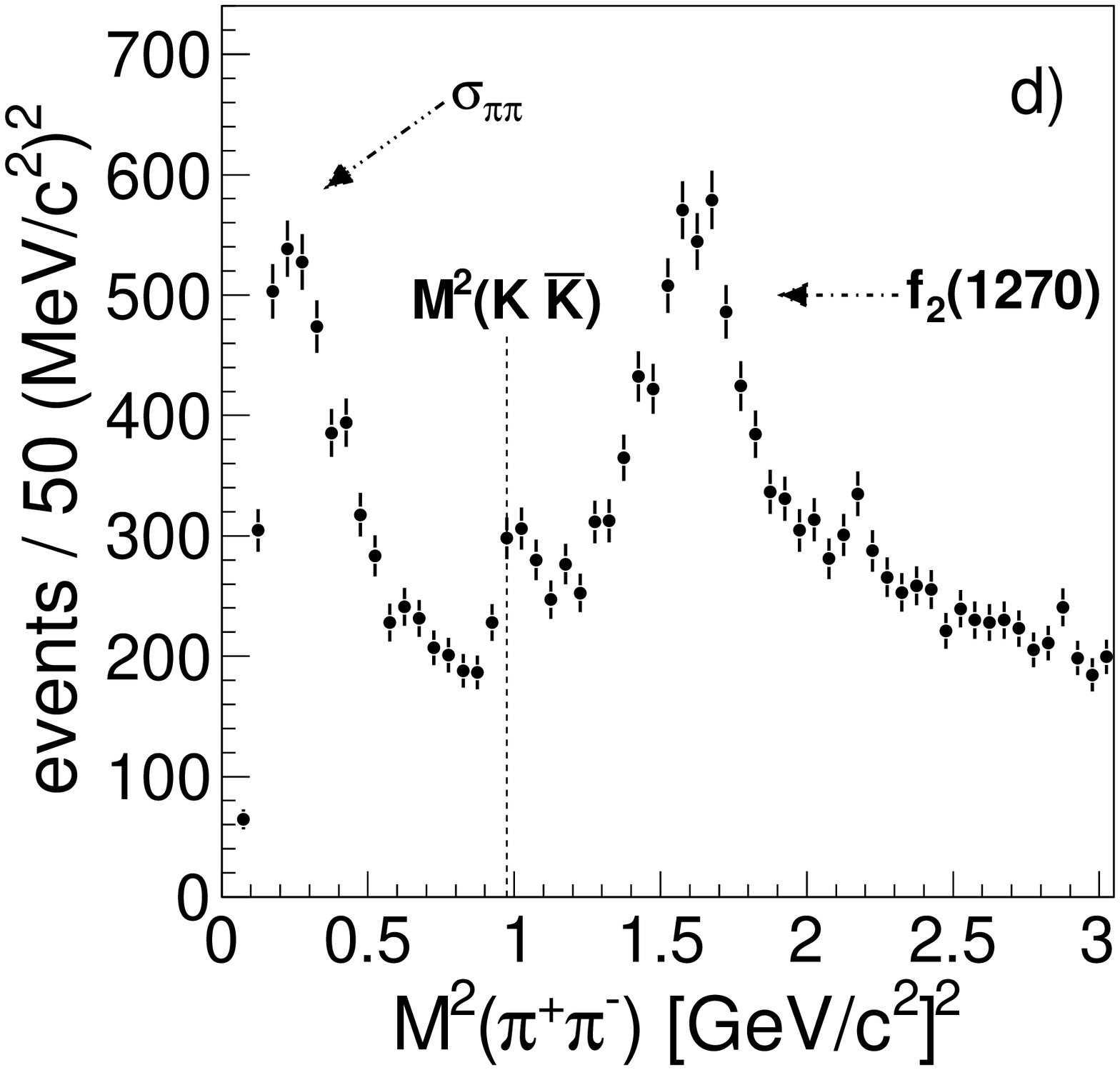,width=7.0cm} 
\caption{\label{fig:Dalitz} Dalitz plots obtained from selected  $\chico$ candidates from (a) data and (b) exclusive MC, 
assuming the $\atwop$ is the only structure produced. The (c) $\eta\pi$ and (d) $\pip\pim$ projections show various structures, 
which can also be identified by arrows in the Dalitz plot (a). 
Vertical dotted lines in plots (c) and (d) indicate the thresholds for producing the $\etap\pi$ or $\KKB$ in the $\eta\pi$ or $\pi\pi$ 
space, respectively.}
\end{figure*}

The Dalitz plot for selected signal events is shown in
Fig.~\ref{fig:Dalitz}(a). Two-body structures reported in previous
analyses of the $\chico\to\eta\pip\pim$ decays, 
by BESII~\cite{BESIIChic} and CLEO~\cite{CLEOChic, CLEOMK}, 
the $\azero\pi$, $\atwo\pi$ and $\ftwo\eta$, are
indicated by the long-dash-dotted, dashed and dash-dotted arrows
pointing into the Dalitz space, respectively.  One feature of this
distribution is the excess of events in the upper left corner of the
Dalitz plot (a), pointed to by the dotted arrows, which cannot be
associated with known structures observed in previous analyses of this
$\chico$ decay.  We hypothesize this is due to $\atwop$
production. The expected Dalitz plot of a $\atwopAmp$ signal is shown
in Fig.~\ref{fig:Dalitz}(b), obtained assuming that the $\atwop$ is
the only structure produced.  The $\atwop\to\eta\pip$ and
$\atwop\to\eta\pim$ components cannot be easily identified along the
dotted arrows in the Dalitz plot, Fig~\ref{fig:Dalitz}(a), but their
crossing in the plot shown in Fig.~\ref{fig:Dalitz}(b) visually
matches the excess of events in the upper left corner of the Dalitz
plot of Fig.~\ref{fig:Dalitz}(a).

The distributions of the square of the invariant mass are shown in
Fig.~\ref{fig:Dalitz}(c) for $\eta\pi$ and (d) for $\pip\pim$.
Structures that correspond to $\azero$, $\atwo$ and $\ftwo$ production
are evident, as well as a low-mass $\pi\pi$ peak, sometimes referred
to as the $\sigma$ state. In each of these two distributions there is
a visible threshold effect. In the $\pi\pi$ distribution, there is a
structure above the $\KKB$ threshold, which is too broad to result
from the $\fzero$ alone. In the $\eta\pi$ distribution, the broadening
of the $\azero$ peak around 1.2~$\gev^{2}/c^4$ could be associated
with the $\etap\pi$ threshold.  By examining various regions in the
Dalitz space, we conclude that the cross-channel contamination, or
reflections, are not associated with these threshold effects in the
data. In order to eliminate background as the source of these peculiar
line shapes, background studies are performed. Namely, we increased the
background level by relaxing the kinematic constraint to
$\chi^2_{NC}/NC < 10$ and also suppressed more background by requiring
$\chi^2_{NC}/NC < 5$. In addition we varied the limits on tagging
$\eta$ and $\chico$ candidates, as explained in Sec.~\ref{SysErr}.

It is possible that the $\pi\pi$ line shape results from a destructive
interference between the $\fzero$ and other components of the $\pi\pi$
$S$-wave. It has been known for some time that the $\azero\to\eta\pi$
line shape is affected by the proximity of the $\KKB$ threshold to the
$\azero$ mass ~\cite{FLATTE76}. If the $\azero\to\etap\pi$ coupling
appears to be important for describing the $\azero\to\eta\pi$
distribution, this would be an example when a virtual channel is
influencing the distribution of another decay channel, despite its
threshold being far away from the resonance peak.  We use an amplitude
analysis (AA), described in the next section, to help in answering the
above questions, and to determine the nature and significance of the
``crossing structure'' discussed.

\section{Amplitude analysis}\label{AmpAna}

To study the substructures observed in the $\chico\to\eta\pip\pim$ decays,
we use the isobar model, in which it is assumed that the decay
proceeds through a sequence of two-body decays, $\chico \to R h_b$;
$R\to h_1 h_2$, where either an isospin-zero ($R\to\pi\pi$) or
isospin-one ($R\to\eta\pi$) resonance is produced, with the total spin
$J$, and relative orbital angular momentum $L$ with respect to the
bachelor meson, $h_b$. For resonances with $J>0$, there are two
possible values of $L$
that satisfy the quantum number conservation for the $1^{++} \to
(\jpc) 0^{-}_L$ transition.

We use the extended maximum likelihood technique to find a set of
amplitudes and their production coefficients that best describe the
data. The method and complete description of amplitudes constructed
using the helicity formalism are given in Ref.~\cite{CLEOMK}, with two
exceptions.

The first difference is that the events from the $\eta$-sidebands are
subtracted in the likelihood function $\mathcal{L}$, with equal
weight given to the left-hand and right-hand sides, using a weighting
factor $\omega = -0.5$. The second difference with respect to
Ref.~\cite{CLEOMK} is that we deviate from the strict isobar model by
allowing production amplitudes to be complex. Isospin symmetry for
$\eta\pi^{\pm}$ resonances is imposed.

In the minimization process of the expression $-2\ln{ \mathcal{L} }$,
the total amplitude intensity, $I({\bf x})$, constructed from the
coherent sum of relevant amplitudes, is bound to the number of
observed $\chico$ candidates by using the integral
\begin{equation}\label{eq:TotI}
	\mathcal{N}_{\chico} = \int \xi(\bf{x}) I(\bf{x}) d \bf{x}, 
\end{equation}
where $\bf{x}$ represents the kinematic phase space, while
$\xi(\bf{x})$ is the acceptance function, with the value of one (zero)
for accepted (rejected) exclusive MC events. The proper normalization
of different $\eta$ channels is ensured by using exclusive MC samples,
generated with sample sizes proportional to the $\eta$ branching
fractions, listed in Table~\ref{tab:fs}. If the complete generated
exclusive MC set is used in the MC integration, then
Eq.~(\ref{eq:TotI}) provides the acceptance corrected number of $\chico$
events, adjusted by subtracted background contributions. In this case, $\xi(\bf{x})
\equiv 1$ for all MC events.  Fractional contributions,
$\mathcal{F_{\alpha}}$, from specific amplitudes, $A_{\alpha}$, are
obtained by restricting the coherent sum in $I(\bf{x})$ to
$I_{\alpha}(\bf{x})$, so that
\begin{equation}\label{eq:FracI}
	\mathcal{F_{\alpha}} = \frac{ \int I_{\alpha}(\bf{x}) d \bf{x} }{ \int I(\bf{x}) d \bf{x} }.
\end{equation}
The numerator represents acceptance-corrected yield of a given
substructure, used to calculate relevant branching fractions,
$\mathcal{B}_{\alpha}$.  Errors are obtained from the covariance
matrix using proper error propagation, so for a given substructure,
the errors on $\mathcal{B}_{\alpha}$ and $\mathcal{F}_{\alpha}$ are not
necessarily the same.

The decay chain $\fulldecaya$ is described by amplitudes constructed
to take into account the spin alignment of the initial
state and the helicity of the radiated photon. 
Linear combinations of helicity amplitudes can be used to construct
amplitudes in the multipole basis, matching the electric dipole ($E1$)
and magnetic quadrupole ($M2$) transitions. The
$\psip\to\gamma\chico$ decay is dominated by the $E1$
transition (CLEO)~\cite{CLEOE1M2}, and a small $M2$ contribution ($\approx
3\%$) can be treated as a systematic uncertainty.

\subsection{\boldmath Mass dependent terms, $T_{\alpha}(s)$}\label{AmpDef}

The dependence of amplitude $A_{\alpha}$ on the energy can be separated from
its angular dependence, employing a general form $p^L q^J T_{\alpha}(s)$,
if the width of the $\chico$ is neglected. Here, $p$ and $q$ are
decay momenta for decays $\chico\to R_J h_b$ and $R_J \to h_1 h_2$ in
the rest frame of the $\chico$ and a resonance $R_J$, respectively,
while $s = m_{12}^2$ is the squared invariant mass of the corresponding isobar 
products ($\pi\pi$ or $\eta\pi$). 
For most resonances, we use relativistic Breit-Wigner (BW)
distributions, with spin-dependent Blatt-Weisskopf
factors~\cite{Blatt51}.
For the $\azero$ and $\pi\pi$ $S$-wave line shapes,
we use different prescriptions explained below.

To account for the nonresonant process $\decaya$, we use an amplitude
constructed as the sum of all possible final state combinations of
helicity amplitudes constrained to have the same production strength,
with no dependence on the invariant mass of the respective two-body
combinations.

\subsubsection{\boldmath Parametrization of $\azero$ \label{AzeAmp} }

\begin{figure*}[htbp]
\centerline{\hbox{ 
\epsfig{file=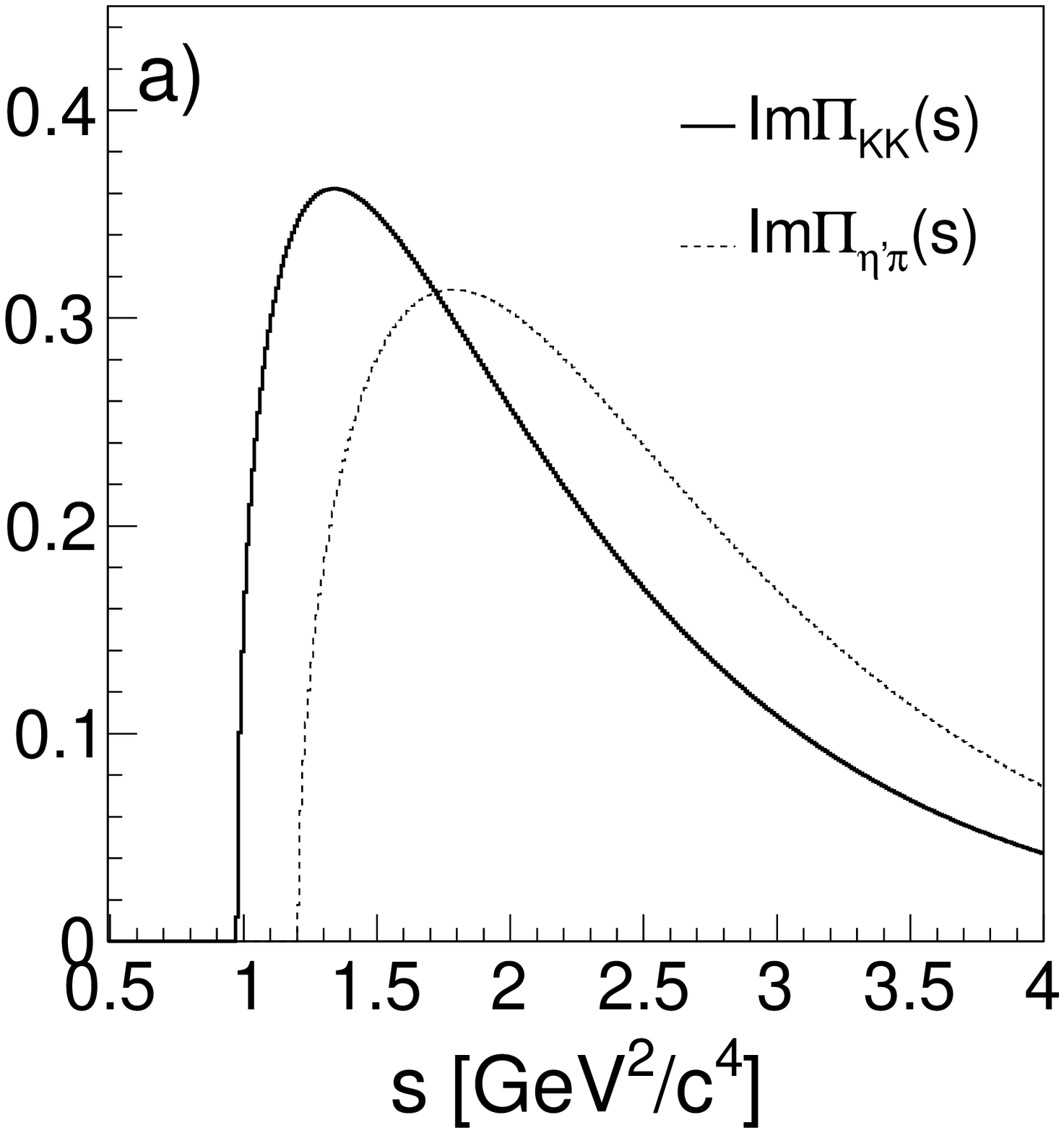,width=6.0cm} 
\epsfig{file=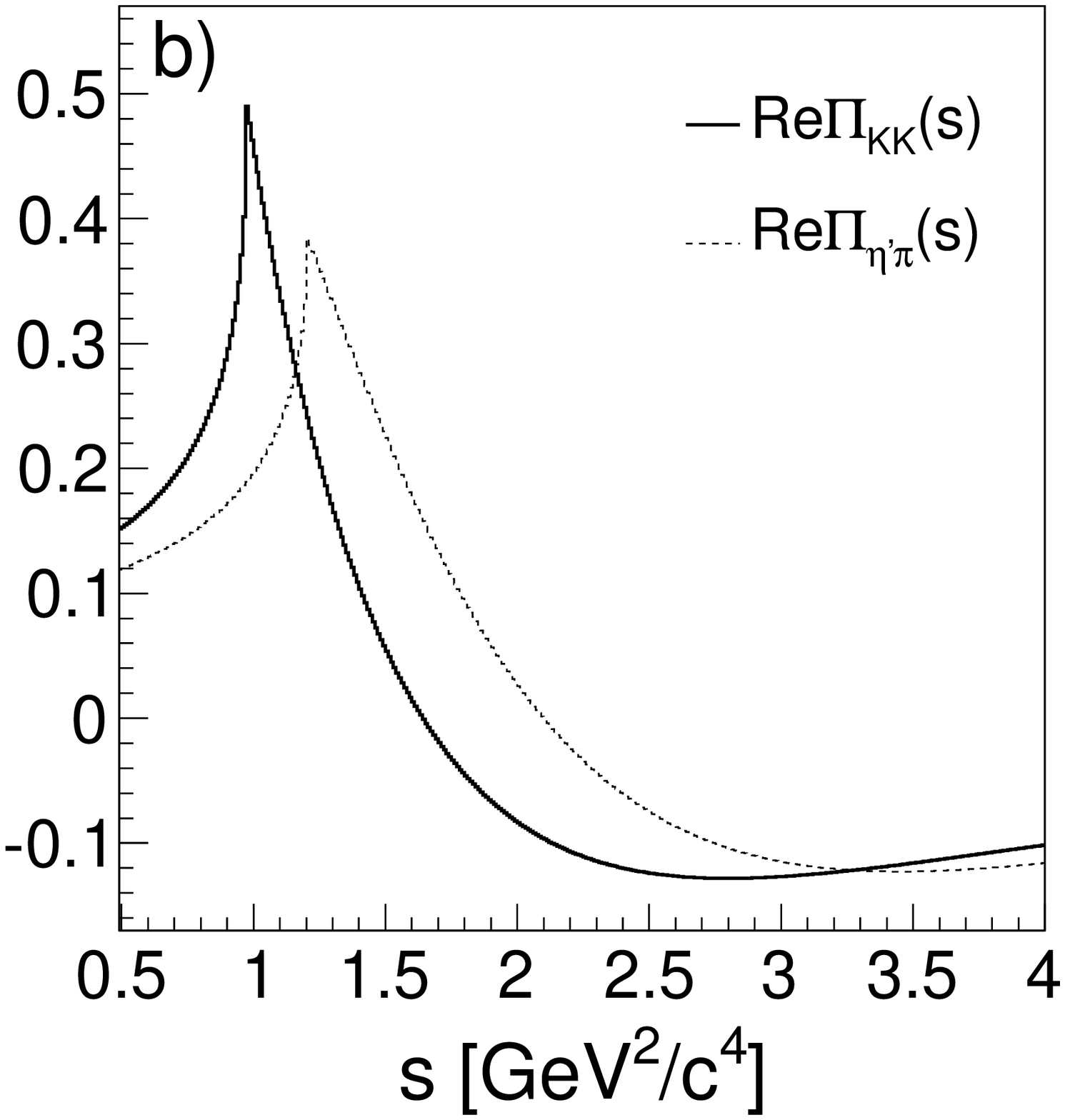,width=6.0cm} 
}}
\caption{
Line shapes of (a) $\text{Im} \Pi(s)$ and (b) $\text{Re} \Pi(s)$ for the $\KKB$ and $\etap\pi$ production with arbitrary normalization.
}\label{fig:DispInt}
\end{figure*}

Instead of using the usual Flatt\'e formula~\cite{FLATTE76} to
describe the $\azero$ line shape, we use dispersion integrals,
following the prescription given in Ref.~\cite{DBUG78_08}.  We
consider three $\azero$ decay channels, the $\eta\pi$, $\KKB$, and
$\etap\pi$, with corresponding coupling constants, $g_{ch}$, and use
an appropriate dispersion relation to avoid the problem of a false
singularity~\cite{AnisSar} present in the $\etap\pi$ mode (see
the discussion at the end of this section).  The $\azero$ amplitude is
constructed using the following denominator:
\begin{equation}\label{eq:a0Disp}
D_{\alpha}(s) = m^2_0 - s - \sum_{ch} \Pi_{ch}(s),
\end{equation} 
where $m_0$ is the $\azero$ mass and $\Pi_{ch}(s)$ in the sum over channels is a complex function, with imaginary part  
\begin{equation}\label{eq:ImPij}
	\text{Im} \Pi_{ch}(s) = g_{ch}^2 \rho_{ch}(s) F_{ch}(s),
\end{equation} 
while real parts are given by principal value integrals,
\begin{equation}\label{eq:RePij}
\text{Re} \Pi_{ch}(s) = \frac{1}{\pi} P \int^{\infty}_{s_{ch}} \frac{\text{Im} \Pi_{ch}(s^{\prime}) ds{^\prime}}{(s{^\prime} - s)}.
\end{equation} 
In the above expressions $\rho_{ch}(s)$ is the available phase space
for a given channel, obtained from the corresponding decay momentum
$q_{ch}(s)$: $\rho_{ch}(s) = 2q_{ch}(s)/\sqrt{s}$. The integral in
Eq.~(\ref{eq:RePij}) is divergent when $s \to \infty$, so the phase
space is modified by a form factor $ F_{ch}(s) = e^{ -\beta
q^2_{ch}(s)} $, where the parameter $\beta$ is related to the
root-mean-square (rms) size of an emitting
source~\cite{DBUG78_08}. We use $\beta = 2.0~[\gevcc]^{-2}$
corresponding to rms~=~0.68~fm, and we verify that our results are not
sensitive to the value of $\beta$.  The integration in
Eq.~(\ref{eq:RePij}) starts from the threshold for a particular
channel, $s_{ch}$, which conveniently solves the problem of the
analytical continuation in special cases of final state configurations
like the $\azero\to\etap\pi$, when the decay momentum below the
threshold ($s < m_{\etap} + m_{\pi}$) becomes real again for $s <m_{\etap} - m_{\pi}$. 
Figure~\ref{fig:DispInt} shows the shapes of
(a) $\text{Im} \Pi_{ch}(s)$ and (b) $\text{Re} \Pi_{ch}(s)$, for the $\KKB$ and
$\etap\pi$ channels, for arbitrary values of the coupling
constants. In the final form, the real parts in the denominator of
Eq.~(\ref{eq:a0Disp}) are adjusted by $\text{Re}\Pi_{ch}(m_0)$ terms:
$\text{Re}\Pi_{ch}(s) \to \text{Re}\Pi_{ch}(s) - \text{Re}\Pi_{ch}(m_0)$.

\subsubsection{\boldmath $\pi\pi$ $S$-wave model}\label{Swave}

The $\pi\pi$ $S$-wave parametrization follows the prescription given in Ref.~\cite{CLEOMK}, in which two independent processes for
producing a $\pi\pi$ pair are considered: direct $(\pi\pi)_S \to (\pi\pi)_S$, and production through kaon loops, $(\KKB)_S \to (\pi\pi)_S$. 
Amplitudes corresponding to these scattering processes, labeled $S_{\pi\pi}(s)$ and $S_{\KKB}(s)$, are based on di-pion phases and intensities 
obtained from scattering data~\cite{LEON99},
which cover the $\pi\pi$ invariant mass region up to 2~$\gevcc$. 
The $\spp(s)$ component is adapted in Ref.~\cite{CLEOMK} to account for differences in 
the $\pi\pi$ production through scattering and decay processes, using the denominator, $D(s)$, extracted from scattering
experiments. The $\spp(s)$ amplitude in this analysis takes the form:
\begin{equation}\label{eq:SppExp}
   \spp(s) = c_0 \szs + \sum_{i=1} c_i z^i_{s_{\KKB}}(s) \szs + \sum_{i=1} c'_i z^i_{s'}(s) \szs.
\end{equation}
The common term in the above expression, $\szs = 1 / D(s)$, is expanded using conformal transformations of the type
\begin{equation}\label{eq:Zth}
z_{s_\text{th}}(s) = \frac{ \sqrt{s + s_0} - \sqrt{ s_\text{th} - s } }{ \sqrt{s + s_0} + \sqrt{ s_\text{th} - s } },
\end{equation}
which is a complex function for $s > s_{th}$. Equation~(\ref{eq:SppExp}) features two threshold functions, $z_{s_\text{th}}(s)$, one corresponds 
to $\KKB$ production with $s_{\KKB} = 4m^2_{K}$, while another with $s_\text{th} = s'$ could be used to examine other possible threshold effects 
in di-pion production. The $c_i$, $i=1,2$ are production coefficients to be determined.

\begin{figure*}[htbp]
\centerline{\hbox{ 
\epsfig{file=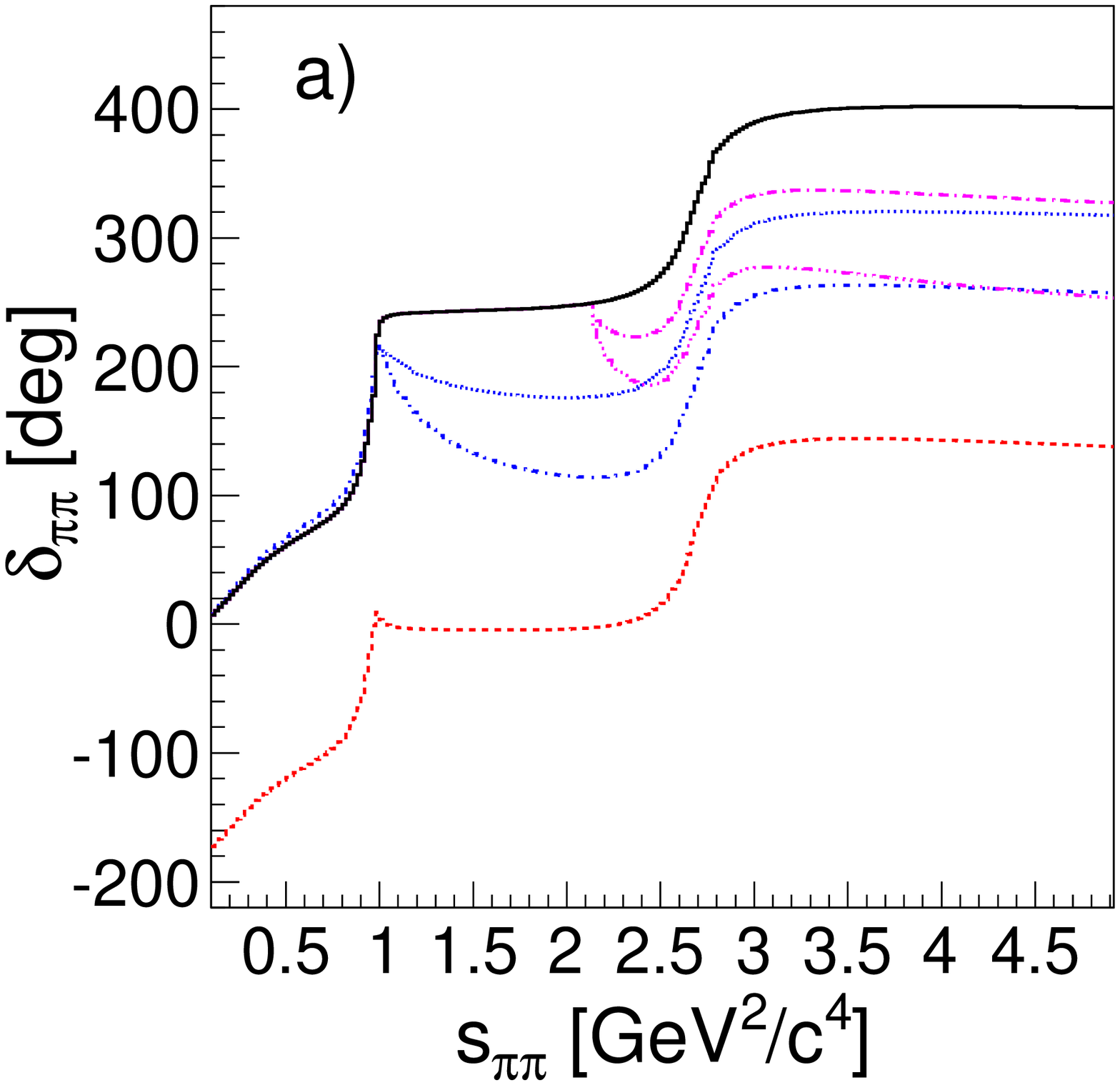,width=7.0cm} 
\epsfig{file=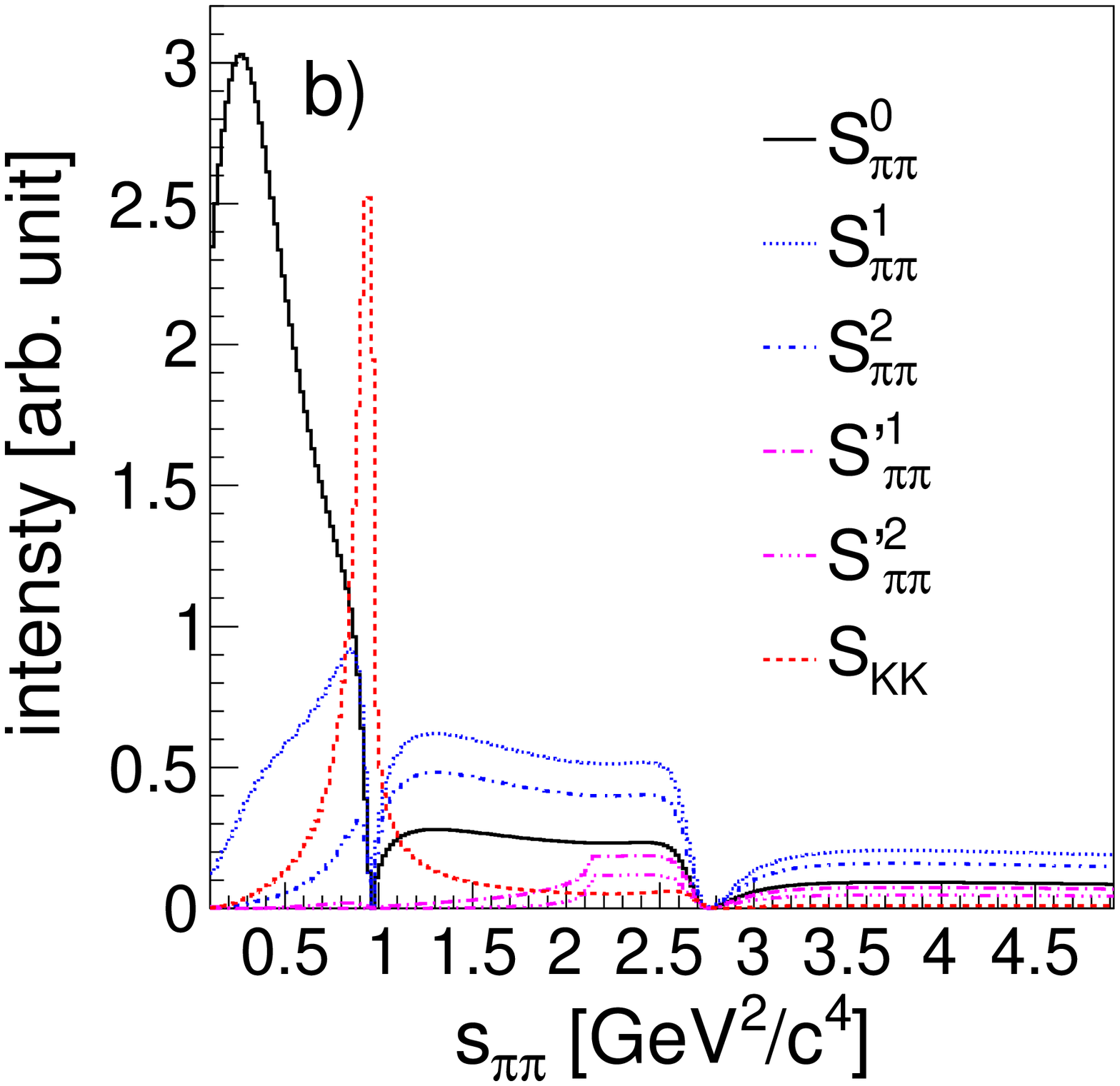,width=7.0cm} 
}}
\caption{The (a) phase and (b) intensity of the $\pi\pi$ $S$-wave
 components. Red (dashed) histograms represent the $\skk$ amplitude, blue
 histograms (doted and dash-doted) are obtained using $S^i_{\pi\pi} = z^i_{\KKB}\sppz$
 terms, while purple (long-dash-doted and dash-three-doted) represent $S'^{i}_{\pi\pi} = z^i_{s'} \sppz$
 terms. } \label{fig:Spp}
\end{figure*}

Figure~\ref{fig:Spp} shows the (a) phase and (b) intensity of various components used in constructing the $\pi\pi$ $S$-wave amplitude based
on two functions given by Eq.~(\ref{eq:Zth}), with different thresholds: $z_{\KKB}(s)$ and $z_{s'}(s)$. The following convention is used:
$S^i_{\pi\pi}(s) = z^i_{\KKB} \szs$, $S'^i_{\pi\pi}(s) = z^i_{s'} \szs$. Components are arbitrarily scaled, and we set 
$\sqrt{s'} \sim 1500~\mevcc$, similarly to the value used later in analysis. The parameter $s_0 = 1.5~(\gevcc)^2$ can be used to adjust the left-hand cut in the complex plane, and the 
same value is used in all components. 

\section{Results}\label{Rez}

We present results from the amplitude analysis of the full decay $\fulldecaya$, reconstructed in three major $\eta$ decay modes. 
The optimal solution to describe the data is found by using amplitudes with fractional contributions larger than 0.5\% and significance
larger than 5$\sigma$. The significance for each amplitude $\alpha$ is determined from the change in likelihood with respect to the
null hypothesis, $ \Delta \Lambda = -2 \ln{\mathcal{L}_0 / \mathcal{L}_{\alpha}}$. The null hypothesis for a given amplitude is found 
by excluding it from the base-line fit, and the corresponding amplitude significance is calculated taking into account the change in
the number of degrees of freedom, which is two (four) for $J=0$ ($J>0$) amplitudes.

\begin{figure*}[htbp]
\centerline{\hbox{ 
\epsfig{file=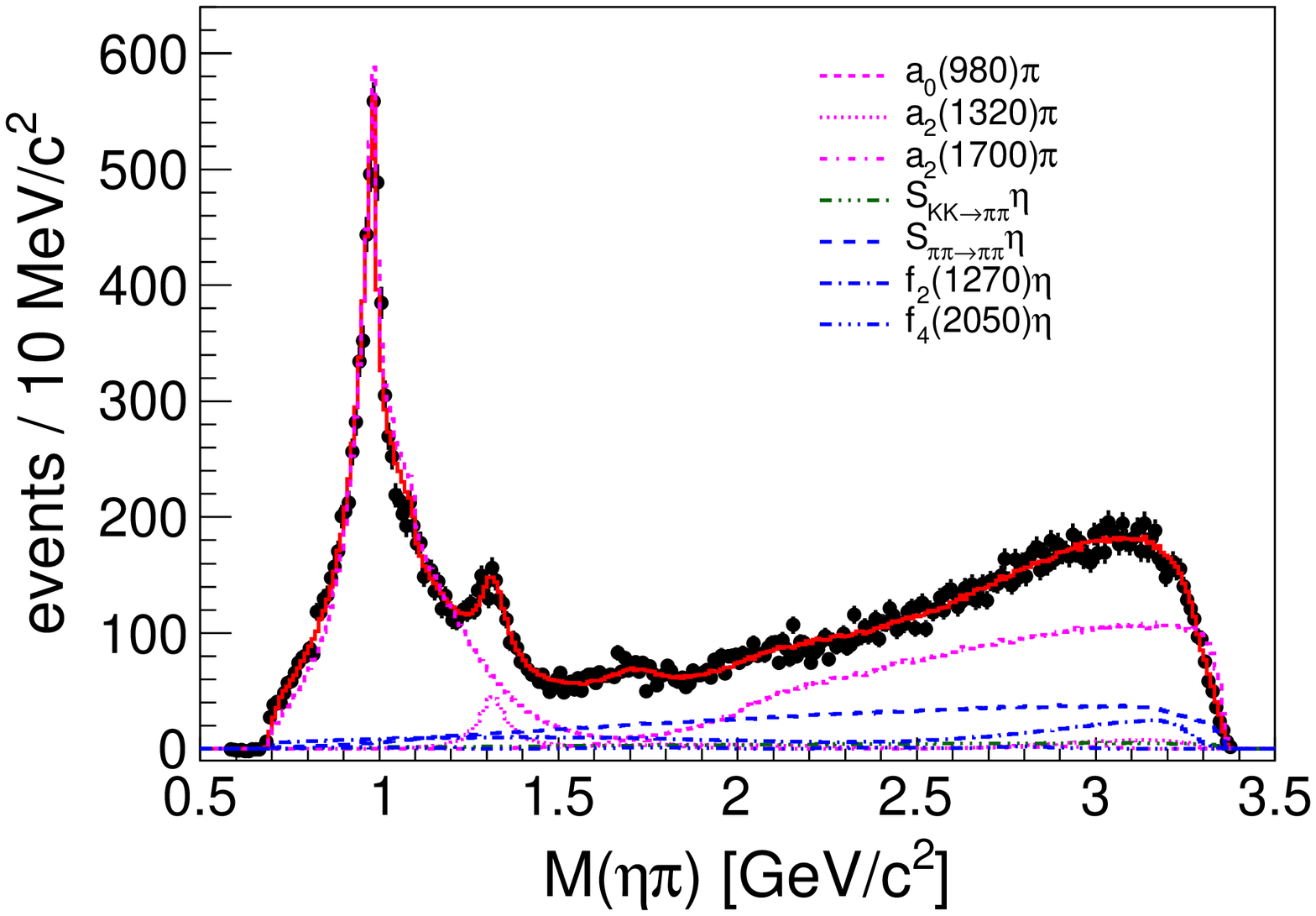,width=7.0cm} 
 \put(-162,120){a)}
\epsfig{file=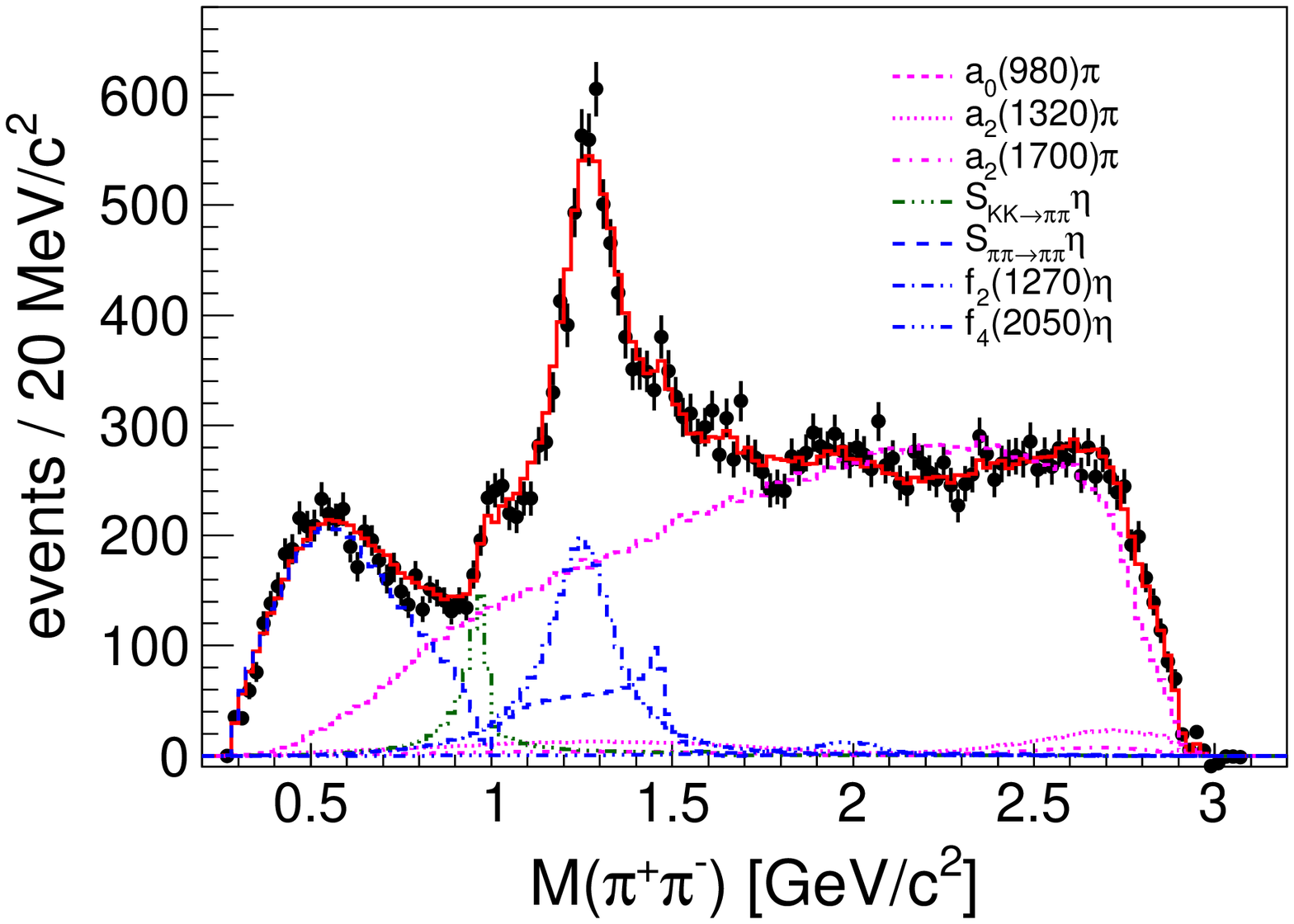,width=7.0cm}  
 \put(-157,120){b)}
}}
\caption{ \label{fig:m12m23} Projections in the (a) $\eta\pi$ and (b)
$\pip\pim$ invariant mass from data, compared with our base-line fit
(solid curve) and corresponding amplitudes (various dashed and dotted
lines).  All features of the data, including structures discussed in
Sec.~\ref{SigAll} are reproduced rather well.  }
\end{figure*}

The most dominant amplitude in this reaction is $\azeAmp$, as
evident from the $\eta\pi$ projection of the Dalitz plot,
Fig.~\ref{fig:Dalitz}(c). Other amplitudes used in our base-line fit
include the $\skkAmp$, $\sppAmp$, $\ftwoAmp$, $\ffoAmp$, $\atwoAmp$
and $\atwopAmp$, where masses and widths of resonances described by BW
functions are taken from the PDG~\cite{PDG14}, while the $\atwop$ and
$\azero$ parameters are free parameters to be determined by the fit in this work. The mass projections are
shown in Fig.~\ref{fig:m12m23}, and the corresponding fractional
contributions and significances are listed in
Table~\ref{tab:Frac-sig}. For amplitudes with spin $J>0$ both orbital
momentum components are included.

The following components form the $\spp(s)$ amplitude:
\begin{equation}\label{eq:SppFin}
	\spp(s) = c_0 \szs + c_{1} S^1_{\pi\pi}(s) + c'_{1} S'^1_{\pi\pi}(s) + c'_{2} S'^2_{\pi\pi}(s).
\end{equation}
As indicated earlier, the threshold used to construct the $S^1(s)$ term is $s_{\KKB} = 4m^2_{K}$. 
The threshold for the $S'^i(s)$ components ($i=1,2$) is $s' = 2.23~[\gevcc]^2$, which is close to the mass of the $f_0(1500)$,
and it is responsible for the peaking of the $\spp\eta$ amplitude in this region, Fig.~\ref{fig:m12m23}(b).  
In fact, the  $S'^i(s)$ components are used instead of the $f_0(1500)\eta$ 
amplitude, which would be needed in the optimal solution if only  
threshold functions $z^i_{\KKB}(s)$  were used in the expansion of the 
$\spp(s)\eta$ amplitude.  
With these additional terms, the contribution and significance of $\pi\pi$ scalars, 
the $f_0(1370)$, $f_0(1500)$ and $f_0(1710)$, is negligible, for each. 
Although this particular set of amplitudes respects the
unitarity of the $\pi\pi$ S-wave, we use the sum of BW to model 
other spins and final states, namely the $\ftwo$, $\ffo$, $\atwo$ and $\atwop$. 
Our approach provides reasonable modeling 
of the $\pi\pi$ line shape, and the sum of all $\pi\pi$ $S$-wave components, $\skk$ and $\spp$, 
is reported in Table~\ref{tab:Frac-sig}.   

Besides the $f_0(1370)$, $f_0(1500)$, and $f_0(1710)$, other conventional
resonances are probed, including the $f_0(1950)$, $f_2(1525)$,
$f_2(2010)$, and $a_0(1450)$,
with parameters fixed to PDG values~\cite{PDG14}. They do not pass the
tests for significance and fractional contribution. The non resonant 
$\chico\to\eta\pip\pim $ production is found to be negligible. The search for possible $1^{-+}$ resonances in the $\eta\pi$ final 
state will be presented below. 

\begin{table*}[htbp]
\begin{center}
\caption{Fractional intensities $\mathcal{F}$, and significances of
amplitudes in the base-line fit, with the first and second errors
being statistical and systematic, respectively.  The third error for
the branching fractions for the $\chico\to\eta\pip\pim$ decay and
decays into significant conventional isobars is external (see text).
For exotic mesons only statistical errors on their fractional
contributions are provided. The upper limits for exotic meson
candidates, which include both statistical and systematic
uncertainties, are at the 90$\%$ confidence level.  The coherent sum
of all $\pi\pi$ $S$-wave components, $(\pip\pim)_{S}\eta$, is included
in this report.  Note, the branching fractions for amplitudes of the
type $A_{\alpha}\eta$, involving isobars decaying into $\pip\pim$, are
the products of $\chico\to A_{\alpha}\eta$ and $A_{\alpha}\to\pip\pim$
rates.  Branching fractions for isobars decaying into $\eta\pi$
include charge conjugates.
\label{tab:Frac-sig}
}
\begin{tabular}{lc@{\hskip 0.2cm}c@{\hskip 0.2cm}c}
\hline	\hline	
Decay &	$\mathcal{F}$	[\%] & Significance [$\sigma$] & $\mathcal{B}(\chico\to\eta\pip\pim)$ [$10^{-3}$] \\
\hline 
 $\eta\pip\pim$	& $\dots$	& $\ldots$ &
 4.67	$\pm$	0.03	$\pm$	0.23	$\pm$	0.16	 \\
\hline 

$a_0(980)^+\pim$ & 72.8 $\pm$ 0.6 $\pm$ 2.3 & $> 100$ &
                3.40	$\pm$	0.03	$\pm$	0.19	$\pm$	0.11 \\
$a_2(1320)^+\pim$	& 3.8 $\pm$ 0.2 $\pm$ 0.3  & 32 &   
		0.18	$\pm$	0.01	$\pm$	0.02	$\pm$	0.01 \\

$a_2(1700)^+\pim$	& 1.0 $\pm$ 0.1 $\pm$ 0.1   & 20 &
		0.047	$\pm$	0.004	$\pm$	0.006	$\pm$	0.002 \\	
 
$S_{\KKB}\eta$	& 2.5 $\pm$ 0.2 $\pm$ 0.3   & 22 &
		0.119	$\pm$	0.007	$\pm$	0.015	$\pm$	0.004	\\
$S_{\pi\pi}\eta$	& 16.4 $\pm$ 0.5 $\pm$ 0.7   & $ > 100$ &
		0.76	$\pm$	0.02	$\pm$	0.05	$\pm$	0.03 \\
$(\pip\pim)_{S}\eta$ & 17.8	$\pm$ 0.5 $\pm$ 0.6  & ... &  
		0.83	$\pm$	0.02	$\pm$	0.05	$\pm$	0.03 \\
$f_2(1270)\eta$	& 7.8 $\pm$ 0.3 $\pm$	1.1  & $ > 100$ & 
		0.36	$\pm$	0.01	$\pm$	0.06	$\pm$	0.01 \\
 
$f_4(2050)\eta$	& 0.6 $\pm$ 0.1 $\pm$	0.2   & 9.8 &
		0.026	$\pm$	0.004	$\pm$	0.008	$\pm$	0.001	\\

\hline
Exotic candidates &  &  &  U.L. [90$\%$ C.L.]  \\

\hline

$\pi_1(1400)^{+}\pim$ & 0.58$\pm$0.20 & 3.5  &
       $<$      0.046	 \\

       $\pi_1(1600)^{+}\pim$ & 0.11$\pm$0.10 & 1.3  &
                               $<$      0.015	 \\

$\pi_1(2015)^{+}\pim$ & 0.06$\pm$0.03 & 2.6 & 
    $<$     0.008 	 \\ 
 
\hline\hline
\end{tabular}
\end{center}
\end{table*}

\subsection{\boldmath The $\atwop$ signature}\label{a21700}

All structures listed in Table~\ref{tab:Frac-sig} have been already
reported in the decay $\decaya$, except the $\atwopAmp$.  Its
fractional contribution is around 1\%, and the significance of each
orbital momentum component is more than $10\sigma$. Detailed
background studies are performed to ensure that the background,
remaining after $\eta$-sideband subtraction, is not affecting the
significance and fractional contribution of the $\atwop$.  Results of
fitting the mass and width of the $a_2(1700)$, shown in
Table~\ref{tab:f2a2_opt}, are consistent with the values listed by the
PDG~\cite{PDG14}. To check how the $\atwop$ parameters and fractional
contributions are affected by the $\ftwo$ and $\atwo$, we also fitted
their masses and widths, which are provided in
Table~\ref{tab:f2a2_opt} with statistical uncertainties only.
The mass (width) of the $\ftwo$ is lower (higher) than its nominal value~\cite{PDG14}, 
maybe because of interference with underlying $\pi\pi$ $S$-wave components or threshold effects, 
other than those for the $\KKB$ or $f_0(1500)$ production.

The systematic uncertainties for the $\atwop$ mass and width are obtained by varying parameters of other 
amplitudes within respective uncertainties listed in Ref.~\cite{PDG14}, 
and taking into account variations listed in Table~\ref{tab:f2a2_opt}. 
The $\azero$ errors are shown in Table~\ref{tab:a0par}. 
Variations in the shape of the $\pi\pi$-S wave amplitude 
are taken into account by changing terms in the expansion, Eq.~(\ref{eq:SppFin}).

\begin{figure*}[htbp]
\centerline{\hbox{ 
\epsfig{file=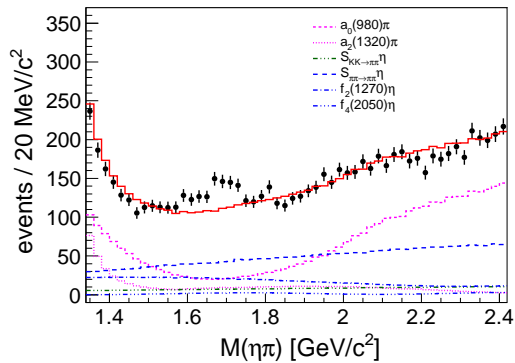,width=7.0cm} 
}}
\caption{ \label{fig:xa21700} 
The $\eta\pi$ invariant mass projection from data in the region (1.3, 2.4)~$\gevcc$, compared with the fit without 
the $\atwop\eta$ amplitude (solid curve). Other amplitudes are plotted (various dashed and dotted lines) for comparison,
while the peak that is associated with the $\atwop$ is evident.  
}
\end{figure*}

We also test the significance of the $a_{2}(1700)$ including
alternative states with the same mass and width, but different spins:
$J=0, 1, 4$.  In all cases, the significance of the $a_2(1700)$ in the
presence of an alternative state exceeds 17$\sigma$. The statistical
significance of the $\atwop$ signal alone is 20$\sigma$.  This result
confirms our hypothesis based on a visual inspection of the Dalitz
plot, Fig.~\ref{fig:Dalitz}(a), that the excess of events in the upper
left corner of the Dalitz space results from the $\atwop$ production,
and it is associated with the crossing of the $a_2(1700)^+\pim$ and
$a_2(1700)^-\pip$ components.
Further, Fig.~\ref{fig:xa21700} shows 
the $\eta\pi$ mass distribution in the region around the expected $\atwop$ peak, 
where data points are compared with a fit when the $\atwop\pi$ amplitude is excluded.

\begin{table*}[htbp]
\caption{\label{tab:f2a2_opt}
The mass and width of the $\atwop$, with statistical and systematic uncertainties. 
Only statistical uncertainties from the $\ftwo$ and $\atwo$ fits are listed.
Comparison with the PDG~\cite{PDG14} values is provided, with all units in $\gevcc$. 
}
\medskip
\begin{center}
\begin{tabular}{l@{\skip 0.2in}c@{\hskip 0.2in}c@{\hskip 0.2in}c@{\hskip 0.2in}c}
\hline
\hline 
&  \multicolumn{2}{c}{BESIII}  & \multicolumn{2}{c}{ PDG~\cite{PDG14} } \\  

Resonance & M  & $\Gamma$ & M & $\Gamma$ \\ 

\hline 
    $a_2(1700)$ & 1.726$\pm$0.012$\pm$0.025 & 0.190$\pm$0.018$\pm$0.030 & 1.732$\pm$0.016 & 0.194$\pm$0.040 \\ %
\hline 

    $f_2(1270)$ & 1.258$\pm$0.003       &  0.206$\pm$0.008    & 1.275$\pm$0.001 & 0.185$\pm$0.003  \\ 

    $a_2(1320)$ & 1.317$\pm$0.002           &  0.090$\pm$0.005     & 1.318$\pm$0.001 & 0.107$\pm$0.005  \\ 

\hline\hline
\end{tabular}
\end{center}
\end{table*}

\subsection{\boldmath $\azero$ parameters}\label{AzePar}

When determining the $\azero$ parameters we use the ratios $R_{21} =
g^2_{\KKB}/g^2_{\eta\pi}$, and $R_{31} =
g^2_{\etap\pi}/g^2_{\eta\pi}$. The resulting values are listed in
Table~\ref{tab:a0par}, where systematic uncertainties  
are obtained by fitting the $\azero$ parameters under different
conditions.  The level of background is varied by changing selection
criteria described in Sec.~\ref{Data}, and by changing the amount of
background subtracted from the $\eta$ sidebands. Effects of the line
shapes of the $\atwo$, $\atwop$, $\ftwo$ and $\ffo$ resonances are
taken into account by varying their masses and widths within the 
respective uncertainties~\cite{PDG14}, and using values from
Table~\ref{tab:f2a2_opt}.  The effect of the $\pi\pi$ $S$-wave shape
is examined in similar way as for the $\atwop$.  The presence of
alternative conventional and exotic resonances is also taken into
account. Our result is not  sensitive to the value of the parameter $\beta$ in Eqs.~(\ref{eq:ImPij}) 
and (\ref{eq:RePij}), within the range of values: $\beta = (2.0\pm 1.0)~[\gevcc]^{2}$. 

For comparison we list two previous results, one from a similar
experiment, CLEO-c, and the other obtained using Crystal Barrel
data. There is a general agreement between different analyses for the
$\azero$ mass and $R_{21}$. The ratio $R_{31}$ was fixed in
Ref.~\cite{DBUG78_08} to the theoretical value provided by
Eq.(\ref{eq:SU3_r31}), while it was consistent with zero in the CLEO-c
analysis, possibly because of smaller statistics. It is not easy to
comment on the difference in values for the $\eta\pi$ coupling, which
could be affected by different normalizations used by different
analyses.

This analysis provides the first nonzero measurement of
the coupling constant $g_{\etap\pi}$. To test the sensitivity of the
$\azero\to\eta\pi$ line shape to the decay $\azero\to\etap\pi$, we repeat
the analysis with $g_{\etap\pi} = 0$, and let the values of the other
parameters free.  The results of this fit are also given in Table~\ref{tab:a0par}.
The likelihood change when the $\etap\pi$ channel is ignored shows
that the significance of a nonzero $g_{\etap\pi}$ measurement is
$8.9\sigma$. The same result is obtained when the analysis is performed
in the presence of the $a_0(1450)$. The values of
the two ratios based on the SU(3) expectation are
\begin{equation}\label{eq:SU3_r21}
 g^2_{\KKB} / g^2_{\eta\pi} = 1/(2\cos^2{\phi}) = 0.886 \pm0.034,
\end{equation}
\begin{equation}\label{eq:SU3_r31}
 g^2_{\etap\pi} / g^2_{\eta\pi} = \tan^2{\phi} = 0.772 \pm0.068,
\end{equation}
which depend on the choice of the $\eta-\etap$ mixing angle; $\phi = (41.3\pm1.2)^{\circ}$ in this case~\cite{DBUG78_08}.  
Our result is consistent with Eq.~(\ref{eq:SU3_r31}) within
1.5$\sigma$, based on the quoted uncertainties.

\begin{table*}[htbp]
\caption{
Parameters of the $a_0(980)$ determined from the fit using the dispersion relation of Eqs.~(\ref{eq:a0Disp}-\ref{eq:RePij}),
compared to results from previous analyses.
Bold values indicate quantities that are fixed in the fit. 
\label{tab:a0par}
}
\begin{center}
\begin{tabular}{lc@{\hskip 0.12in}c@{\hskip 0.12in}c@{\hskip 0.12in}c}
\hline\hline	
Data  & $m_0$ [$\gevcc$] & $g^2_{\eta\pi}$ $[\gevcc]^2$ & $g^2_{\KKB}/g^2_{\eta\pi}$ & $g^2_{\eta'\pi}/g^2_{\eta\pi}$ \\ 
\hline 

    CLEO-c~\cite{CLEOMK} & $0.998\pm0.016$ & $0.36\pm0.04$ & $0.872\pm0.148$ & $0.00\pm0.17$  \\   
C.Barrel~\cite{DBUG78_08} & $0.987\pm0.004$ & $0.164\pm0.011$ & $1.05\pm0.09$ & \bf{0.772}  \\ 
    BESIII & 0.996$\pm$0.002$\pm$0.007 & 0.368$\pm$0.003$\pm$0.013 &  0.931$\pm$0.028$\pm$0.090 & 0.489$\pm$0.046$\pm$0.103  \\  
\hline

BESIII ($R^2_{31}\equiv0$) &  0.990$\pm$0.001 & 0.341$\pm$0.004  &  0.892$\pm$0.022 & \bf{0.0}   \\ 

\hline\hline
\end{tabular}
\end{center}
\end{table*}

\subsection{\boldmath Search for $\eta\pi$ $P$-wave states}\label{PiOne}

We examine possible exotic meson production in the $\eta\pi$ invariant 
mass region from 1.4 to 2.0~$\gevcc$.  Table~\ref{tab:Frac-sig} lists fractional
contributions and significances of three $\jpc=1^{-+}$
candidates, added one at the time to our nominal fit.
Two possible orbital-momentum configurations for an exotic amplitude
are the $S$-wave and $D$-wave, and the significance of each is tested
individually. 
We find that the significance of the $S$-wave is marginal, less than
2$\sigma$ for every $\pi_1$, and the reported significances in
Table~\ref{tab:Frac-sig} result from using the S and D waves together
in the fit. The most significant of the three possible exotic states is
the $\pi_1(1400)$, with a significance of $3.5\sigma$ and fractional
contribution less than $0.6\%$. 
This represents a weak evidence for the existence of the $\pi_1(1400)$ because in alternative 
amplitude configurations, when parameters of other amplitudes are varied, the significance of this 
state becomes $< 3\sigma$. In the nominal amplitude configuration, the significance of each $\pi_1(1400)$ 
component is less than $3\sigma$, and when taken together, the contribution of the $S$-wave is much smaller 
than the $D$-wave contribution, pointing that the evidence for the $\pi_1(1400)$ is circumstantial.   

Masses and widths of the three exotic
candidates are not very well constrained by previous analyses, and we
vary the respective parameters within listed limits~\cite{PDG14}. Our
conclusion is that there is no significant evidence for an exotic
$\eta\pi$ structure in the $\chico\to\eta\pip\pim$ decays, and we
determine upper limits at the 90\% confidence level 
for the production of each $\pi_1$ candidate.

\subsection{Branching fractions}\label{BFrac}

The branching fraction for the $\chico\to\eta\pip\pim$ decay is given by
\begin{equation}\label{eq:BF}
	\mathcal{B}(\chico\to\eta\pip\pim) = \frac{ \mathcal{P} * \mathcal{N}_{\chico\to\eta\pip\pim}}{N_{\psip} B_{\psip\to\gamma\chico} B_{\eta} \epsilon},
\end{equation}
where the branching fractions $B_{\psip\to\gamma\chico}$ and $B_{\eta}$ are from Ref.~\cite{PDG14};
the latter is listed in Table~\ref{tab:fs}.  
The number of $\psip$, $N_{\psip}$,~\cite{BESnPsip,Psip2012} is provided in Sec~\ref{Data}.
The signal purity, $\mathcal{P}$, given in Sec.~\ref{BcgSup}, takes into account that 
the number of $\chico$ obtained from the amplitude analysis 
includes the background not accounted for by the sideband subtraction.  
Using Eq.~(\ref{eq:TotI}) we obtain $\mathcal N_{\chico} = 192658 \pm 1075$, 
where the error is from the covariance matrix.
The efficiency in Eq.~(\ref{eq:BF}) is $\epsilon \equiv 1$, by construction.

Table~\ref{tab:Frac-sig} lists the branching fraction for the
$\decaya$, and branching fractions for subsequent resonance production
in respective isospin states, $\eta\pi^{\pm}$ or $\pip\pim$, where the
first and second errors are statistical and systematic, respectively.
The branching fraction for a given substructure is effectively a
product:
\begin{equation}\label{eq:Bfrac}
\mathcal{B}_{\alpha} = \mathcal{F}_{\alpha} \times \mathcal{B}(\decaya),
\end{equation}
obtained using generated exclusive MC in accordance with
Eq.~(\ref{eq:FracI}). The third error is external, associated with
uncertainties in the branching fractions for the radiative transition
$\psip\to\gamma\chico$ and $\eta$ decays. We also show the total
$\pip\pim$ $S$-wave contribution, obtained from the coherent sum of
the $\skk$ and $\spp$ components. Statistical errors, as well as
systematic ones, for a given fractional contribution and branching fraction
differ, because common systematic uncertainties for all amplitudes
cancel when fractions are calculated, which will be discussed below.

The upper limits for the production of the $\pi_1(1400)\pi$,
$\pi_1(1600)\pi$, and $\pi_1(2015)\pi$ are shown in
Table~\ref{tab:Frac-sig}. The limits are determined by including the
corresponding amplitude in the nominal fit, one at a time. The
analysis is repeated by changing other amplitude line shapes, and the
background level, in a similar fashion used for determining systematic
uncertainties of nominal amplitudes (see Sec.~\ref{SysErr}).  Masses and
widths of exotic candidates are also varied within limits provided by
the PDG~\cite{PDG14}.  The largest positive deviation of the exotic
candidate yield with respect to the corresponding yield from the
modified nominal fit is effectively treated as the systematic
error, summed in quadrature with the statistical error on a given 
exotic state yield.  The resulting uncertainty is used to determine the
90\% confidence level deviation, and added to the "nominal" yield of
an exotic candidate to obtain the corresponding upper limit for the
branching fraction $B(\chico\to\pi_1^+ \pi^-)$.

The branching fractions for the substructures in $\decaya$ decays
reported by the PDG~\cite{PDG14} are compared in
Table~\ref{tab:bf_cmp} with the values measured in this work, and with
the previous most precise measurement (CLEO-c)~\cite{CLEOMK}.  The
measurement for the $\ftwo$ production is adjusted to account for the
measured relative $\ftwo\to\pip\pim$ width.  There is a rather large
discrepancy between the values for the two most dominant substructures
listed by the PDG and the two most recent measurements.  There is very
good agreement between the last two measurements, suggesting that the
PDG values on two-body structures observed in $\decaya$ need to be
updated.

\begin{table}[htbp]
\caption{\label{tab:bf_cmp}
Comparison between recent measurements of the branching fractions $B(\chico\to\eta\pip\pim)$, 
and with the PDG values.
}
\begin{center}
\begin{tabular}{l@{\hskip 0.3cm}c@{\hskip 0.3cm}c@{\hskip 0.3cm}c} 
\hline
\hline
  \multicolumn{4}{c}{ $\mathcal{B}(\chico\to\eta\pip\pim)$ $\times [10^{-3}]$}  \\
\hline
Decay & BESIII &  CLEO-c~\cite{CLEOMK} & PDG~\cite{PDG14} \\ 
\hline
$\eta\pip\pim$ & 4.67 $\pm$ 0.28 &  4.97 $\pm$ 0.31 &  4.9  $\pm$ 0.5  \\ 
\hline
 $a_0(980)^{+}\pi^{-}$ &  3.40 $\pm$ 0.23 & 3.29 $\pm$ 0.22 & 1.8 $\pm$ 0.6  \\ 
  $f_2(1270)\eta$ &  0.64  $\pm$ 0.11 & 0.66 $\pm$ 0.11 & 2.7 $\pm$ 0.8  \\ 

\hline\hline
\end{tabular}
\end{center}
\end{table}

\section{Systematic Uncertainties}\label{SysErr}

Tables~\ref{tab:bf_sys} summarizes various contributions to the systematic
uncertainties in determining the $\chico\to\eta\pip\pim$ branching
fraction, and Table~\ref{tab:frac_sys} shows the systematics on the fractional
contributions of amplitudes in the nominal fit.
Systematic uncertainties in determining the $\chico\to\eta\pip\pim$ branching
fraction stem from uncertainties in charged track and shower
reconstruction efficiencies, the contribution of the $M2$ multipole
transition, amplitude modeling, the background contribution, and the
uncertainty in the number of $\psip$ produced at
BESIII~\cite{BESnPsip,Psip2012}.  External sources of uncertainty
include the branching fraction $B(\psip\to\gamma\chico)$ and the
fraction of $\eta$ decays, $B(\eta)$ in Eq.~(\ref{eq:BF}).  The external
error affects only branching fractions, not fractional
contributions, 
and it is reported as a separate uncertainty.

Systematic uncertainties associated with the tracking efficiency and shower
reconstruction are 
1\% per track and 1\% per photon. Because of different final states
used in this analysis, tracking and photon uncertainties are weighted
according to the product of branching fractions and efficiencies of
the different
$\eta$ channels, as listed in Table~\ref{tab:fs}.  The resulting systematic
uncertainties for charged tracks and photons are 2.47\% and 3.92\%,
respectively.

The electromagnetic transition $\psip\to\gamma\chico$ is dominated by
the $E1$ multipole amplitude with a small fraction of the $M2$
transition~\cite{CLEOE1M2}. The nominal fit takes only the $E1$ multipole
amplitude.  Adding a small contribution of the $M2$ helicity
amplitude, of $2.9\%$, we find a difference in the branching fraction
of 0.62\%.  This is taken as a systematic uncertainty.

When considering the effects of modeling line shapes of different
amplitudes, we repeat the analysis changing the mass and width of
resonances, $\atwo$, $\ftwo$, and $\ffo$, within respective uncertainties,
and change the $\azero$ and $a_2(1700)$ parameters within the limits
of their statistical uncertainties, given in Tables~\ref{tab:a0par}
and~\ref{tab:f2a2_opt}. We also change BW line shapes by replacing
spin-dependent widths with fixed widths, and take into account the
$\chico$ width and centrifugal barrier as another systematic
error. The largest effect from all these sources is taken as a
systematic uncertainty for the branching fractions and fractional
contributions.

The effect of background is estimated by varying the kinematic-constraint
requirement, changing limits on tagging $\eta$ and $\chico$
candidates, changing the level of suppression of the $\jpsi$ and
$\pio$ productions, and the level of background subtraction. As a
general rule, selection criteria were changed to allow for $\approx
1\sigma$ additional background events, based on the numbers from
the inclusive MC. We use $\chi^2_{NC}/NC < 9$ in all three modes when
varying the kinematic constraint. 
Based on these variations, we conclude that the systematic uncertainty associated 
with the assumption that all charged tracks are pions is negligible. 
To select $\chico$ candidates, we
use photon energy ranges of (0.152--0.187)~$\gev$, in the
$\eta\to\gamma\gamma$ channel, and (0.150--0.190)~$\gev$, in two
$\eta\to 3\pi$ channels. The mass window for the $\eta$ selection is
changed to (0.530--0.565)~$\gevcc$. The $\pio$ suppression window is
reduced to (0.120--0.150)~$\gevcc$ and the $\jpsi$ suppression is
reduced by vetoing two-photon energy within (0.525--0.595)~$\gev$. We
also determine the branching fractions without background subtraction
from $\eta$-sidebands, and the largest effect is listed in
Tables~\ref{tab:bf_sys} and~\ref{tab:frac_sys}.

Some uncertainties that are common for all amplitudes, like tracking, shower
reconstruction, and $N_{\psip}$ errors, cancel out in the fractional
contributions.  However, they are taken into account when branching
fractions are determined.

\begin{table}[htbp]
\caption{\label{tab:bf_sys}
Systematic uncertainties in determining 
the branching fraction $B(\chico\to\eta\pip\pim)$.
The systematic uncertainty per track is $1.0\%$, and for photons it is $1.0\%$ per shower.
}
\begin{center}

\begin{tabular}{lc}																				
\hline\hline																				
Contribution	&	Relative uncertainty ($\%$)	\\ 
MDC tracking 
&	2.5	\\ 
Photon detection
&	3.9	\\ 
 $M2/E1$ 		&	0.6	\\ 
Background		&	1.6	\\ 
Amplitude modeling	&	0.1	\\ 
$N_{\psip}$ 		&	0.7	\\ 
\hline									
Total			&	5.0	\\ 
\hline							
External		&	3.4	\\ 
\hline\hline
\end{tabular}
\end{center}
\end{table}

\begin{table}[htbp]
\caption{\label{tab:frac_sys}
Systematic uncertainties in fractional contributions, in percent, 
for the base-line amplitudes used to model the 
$\chico\to\eta\pip\pim$ decays.
}

\begin{center}
\begin{tabular}{l@{\hskip 0.4in}c@{\hskip 0.2in}c@{\hskip 0.2in}c@{\hskip 0.2in}c}
\hline\hline

Source	&	$M2/E1$	 &	Background  &	$T_{\alpha}(s)$ &		Total \\ 
\hline	
$a_0(980)\pi$	& 0.2	& 0.5	& 3.1	 & 3.2  \\ 
$a_2(1320)\pi$	& 0.5 	& 5.6	& 5.6 	 & 7.9  \\ 
$a_2(1700)\pi$	& 1.4 & 3.8 &	12	 & 13  \\ 
$S_{kk}\eta$	& 3.7 & 2.2 &	11 	 & 11.5	  \\ 
$S_{pp}\eta$	& 1.1 & 1.1	& 4.3 	 & 4.6 	  \\ 
$\pi\pi_{S}\eta$ & 1.5 & 1.1	& 3.0 	 & 3.6	  \\ 
$f_2(1270)\eta$	& 0.5 & 2.3 &	14	 & 15	  \\ 
$f_4(2050)\eta$	& 5.6 & 25 & 18	 & 32	  \\ 
			
\hline\hline
\end{tabular}
\end{center}
\end{table}

\section{Summary}

We analyze the world's largest $\chico\to\eta\pip\pim$ sample, selected with 
very high purity, and find a very prominent $\azero$ peak in the $\eta\pi^{\pm}$
invariant mass distribution.  
An amplitude analysis of the $\fulldecaya$ decay is performed, 
and the parameters of the $\azero$ are determined using a dispersion relation. 
The $\azero$ line shape in its $\eta\pi$ final state 
appears to be sensitive to the details of the $\azero\to\etap\pi$ production, 
and for the first time, a significant nonzero coupling of the $\azero$ to the $\etap\pi$ 
mode is measured with a statistical significance of $8.9\sigma$. 

We also report $a_2(1700)\pi$ production in the
$\chico\to\eta\pip\pim$ decays for the first time, with the mass and
width in agreement with world average values, and this analysis
provides both qualitative and quantitative evidence for the existence
of the $\atwop$. First, the signature of the $\atwop$ in the Dalitz
space is consistent with the observed Dalitz plot distribution. Second,
the $\atwop$ significance from the amplitude analysis is larger than $17\sigma$,
compared to alternative spin assignments, even though the fractional
yield of the $\atwop\pi$ is only 1\%.
This may help in listing the $\atwop$ as an established resonance by the
the PDG~\cite{PDG14}.

We examine the production of exotic mesons that might be expected in
the $\chico\to\eta\pi\pi$ decays: the $\pionef$, $\pi_1(1600)$ and
$\pi_1(2015)$. There is only weak evidence for the $\pionef$ while
other exotic candidates are not significant, and we
determine the upper limits on the respective branching fractions.

\acknowledgments

The BESIII collaboration thanks the staff of BEPCII and the IHEP computing center for their strong support. This work is supported in part by National Key Basic Research Program of China under Contract No. 2015CB856700; National Natural Science Foundation of China (NSFC) under Contracts Nos. 11235011, 11322544, 11335008, 11425524; the Chinese Academy of Sciences (CAS) Large-Scale Scientific Facility Program; the CAS Center for Excellence in Particle Physics (CCEPP); the Collaborative Innovation Center for Particles and Interactions (CICPI); Joint Large-Scale Scientific Facility Funds of the NSFC and CAS under Contracts Nos. U1232201, U1332201; CAS under Contracts Nos. KJCX2-YW-N29, KJCX2-YW-N45; 100 Talents Program of CAS; National 1000 Talents Program of China; INPAC and Shanghai Key Laboratory for Particle Physics and Cosmology; Istituto Nazionale di Sic Nucleare, Italy; Joint Large-Scale Scientific Facility Funds of the NSFC and CAS under Contract No. U1532257; Joint Large-Scale Scientific Facility Funds of the NSFC and CAS under Contract No. U1532258; Koninklijke Nederlandse Akademie van Wetenschappen (KNAW) under Contract No. 530-4CDP03; Ministry of Development of Turkey under Contract No. DPT2006K-120470; The Swedish Resarch Council; U. S. Department of Energy under Contracts Nos. DE-FG02-05ER41374, DE-SC-0010504, DE-SC0012069; U.S. National Science Foundation; University of Groningen (RuG) and the Helmholtzzentrum fuer Schwerionenforschung GmbH (GSI), Darmstadt; WCU Program of National Research Foundation of Korea under Contract No. R32-2008-000-10155-0.


\end{document}

%% file: authors_PUB153.tex
\author{%
    M.~Ablikim$^{1}$, M.~N.~Achasov$^{9,e}$, S. ~Ahmed$^{14}$,
    X.~C.~Ai$^{1}$, O.~Albayrak$^{5}$, M.~Albrecht$^{4}$,
    D.~J.~Ambrose$^{44}$, A.~Amoroso$^{49A,49C}$, F.~F.~An$^{1}$,
    Q.~An$^{46,a}$
% ...
    \\
    \vspace{0.2cm}
    (BESIII Collaboration)\\
    \vspace{0.2cm} {\it
      $^{1}$ Institute of High Energy Physics, Beijing 100049, People's Republic of China\\
% ...
$^{4}$ Bochum Ruhr-University, D-44780 Bochum, Germany\\
      $^{5}$ Carnegie Mellon University, Pittsburgh, Pennsylvania 15213, USA\\
% ...
$^{9}$ G.I. Budker Institute of Nuclear Physics SB RAS (BINP), Novosibirsk 630090, Russia\\
% ...
$^{14}$ Helmholtz Institute Mainz, Johann-Joachim-Becher-Weg 45, D-55099 Mainz, Germany\\
% ...
$^{44}$ University of Rochester, Rochester, New York 14627, USA\\
% ...
$^{46}$ University of Science and Technology of China, Hefei 230026, People's Republic of China\\
% ...
$^{49}$ (A)University of Turin, I-10125, Turin, Italy; (B)University of Eastern Piedmont, I-15121, Alessandria, Italy; (C)INFN, I-10125, Turin, Italy\\
% ...
$^{a}$ Also at State Key Laboratory of Particle Detection and Electronics, Beijing 100049, Hefei 230026, People's Republic of China\\
% ...
$^{e}$ Also at the Novosibirsk State University, Novosibirsk, 630090, Russia\\
% ...
    }
}
   \vspace{0.4cm}


%% file: draft.bbl
\begin{thebibliography}{99}
\bibitem{Psip2012}
Using the same method as in Ref~\cite{BESnPsip}, the number of $\psip$ from the
  2012 run period is found to be $(341.1\pm2.0)\times 10^6$, in preparation for
  publication.
\end{thebibliography}
